\documentclass[aps,prd,twocolumn,superscriptaddress]{revtex4-2}

\usepackage{makecell} % Para forzar saltos de línea en las celdas
\usepackage{graphicx}% Include figure files
\usepackage{dcolumn}% Align table columns on decimal point
\usepackage{bm}% bold math
\usepackage[hidelinks]{hyperref} % Para crear enlaces
\usepackage{xcolor}%Paquete para el color
\usepackage{array} %Para las tablas
\usepackage{booktabs}%Otra para tablas
\usepackage{subfig} % compatible con revtex

\usepackage{enumitem}
\usepackage{soul}%Resaltar

\usepackage{amsmath,amssymb,amsfonts}
\usepackage{mathtools}  % mejoras a amsmath
\usepackage{bm}         % bold math
\usepackage{physics}    % \dd, \pdv, \qty, \abs, \norm, etc. (muy práctico)
\usepackage{tensor}     % índices tensoriales
\usepackage{braket}     % opcional: \bra \ket (si lo usas)
\usepackage{slashed}    % opcional: Dirac slash
\usepackage{mathrsfs}   % \mathscr

\usepackage{amsthm}

\newtheorem{definition}{Definition}[section]
\newtheorem{proposition}{Proposition}[section]
\newtheorem{conjecture}{Conjecture}[section]
\newtheorem{remark}{Remark}[section]
\newtheorem{lemma}{Lemma}[section]

\newtheorem{corollary}{Corollary}[section]

\usepackage{graphicx}
\usepackage{subcaption} % subfiguras modernas
\usepackage{float}      % [H] si lo necesitas (úsalo con cuidado)
\usepackage{caption}

\begin{document}

%\title{Is the existence of a dressed singularity by throat of a wormhole feasible?}
\title{The spacetime structure of an untouchable naked singularity in superstrings theory}

\author{Leonel Bixano}
    \email{Contact author: leonel.delacruz@cinvestav.mx}
\author{Tonatiuh Matos}%
 \email{Contact author: tonatiuh.matos@cinvestav.mx}
\affiliation{Departamento de F\'{\i}sica, Centro de Investigaci\'on y de Estudios Avanzados del Intituto Politécnico Nacional, Av. Intituto Politécnico Nacional 2508, San Pedro Zacatenco, M\'exico 07360, CDMX.
}%

%\affiliation{Departamento de F\'{\i}sica, Centro de Investigaci\'on y de Estudios Avanzados del IPN, Av. I.P.N. 2508, San Pedro Zacatenco, M\'exico 07360, CDMX.}
\date{\today}

\begin{abstract}
According to the Cosmic Censorship Conjecture, naked singularities are believed to be forbidden in nature and must remain hidden by an event horizon. In this work, we present the causal structure of an exact solution to the Einstein-Maxwell-Dilaton equations with five parameters: mass, angular momentum, electric and magnetic charges, and a scale, satisfying constraint equations. For one of the constraints, the solution represents a wormhole (WH), and for the other, a black hole (BH), both with an untouchable ring singularity causally disconnected from the rest of the universe. After topologically defining the concept of Wormhole Cosmic Censorship (WCC), we analyze its metric functions in Papapetrou coordinates to verify metric analyticity in spacetime, construct the Carter-Penrose diagram, and use Boyer-Linquist coordinates to visualize the cladding of the ring singularity by the throat. We conclude that the ring singularity in this WH is clad by the throat, similarly to how the event horizon clads the ring singularity in the Kerr-Newman black hole, thus satisfying the WCC Conjecture. In this work, we show that the topology of the WH throat is such that the two sides of the throat are separated by the singularity but topologically identified, resulting in an instantaneous connection between these two regions. These results are applicable to various theories, including Kaluza-Klein and superstring theory. We provide a rigorous proof that, in the black hole case, the domain of outer communication includes a chronology-violating region and thus supports the existence of closed timelike curves outside the event horizon.
\end{abstract}

\maketitle

 %------------------------------------------------------------
\section{Introduction}
%{\it Introduction}. 
One of the most interesting and surprising predictions of Einstein's equations is, without a doubt, the existence of black holes and singularities in space-time. In 1964 \cite{Penrose:1964wq}, using trapped surfaces and under certain reasonable energy conditions, Penrose demonstrated that singularities in space-time could be generated by gravitational collapse. Around the same time, Hawking examined singularities in cosmology \cite{Hawking:1966sx}, \cite{Hawking:1966jv}, \cite{Hawking:1967ju}. In 1969, Penrose proposed that gravitational collapse singularities are obscured by event horizons \cite{Penrose:1969pc}, excluding the presence of naked singularities, this conclusion is called the cosmic censorship conjecture, which was later refined by Hawking and Ellis \cite{Hawking:1973uf}. Despite its logic, the conjecture remains unproven. However, the works \cite{Eardley:1978tr} and \cite{Shapiro:1991zza} suggest by numerical simulations that naked singularities occur in some collapses, challenging this conjecture. The studies \cite{Joshi:2000fk} and \cite{Joshi:2011qq} explore in more detail the existence and stability of naked singularities and possible violations of the conjecture.

In a previous work \cite{Matos:2012gj} the concept of Wormhole (WH) Cosmic Censorship is introduced, similar to the Penrose hypothesis, but for WHs. The study \cite{Miranda:2013gqa} showed numerically that a null geodesic cannot reach the ring singularity due to the infinite potential and \cite{DelAguila:2018gni} provides an analytical proof of the causal disconnection of the ring singularity assuming slow rotation, but its causal structure and Penrose diagram remained challenging.
In this work, we construct the Penrose diagram and analyze the causal structure of a WH, derived from an exact Einstein-Maxwell-Dilaton solution, and show the causal structure of the WH. We demonstrate that the WH topology is complex and, for the first time, show that the Penrose diagram shows that the ring singularity is causally disconnected from the rest of the universe, giving rise to the Wormhole Cosmic Censorship Conjecture (WCCC). 

The Einstein-Maxwell-Dilaton Lagrangian is given by
\begin{equation}\label{LagrangianoTesisUnidades}
    \mathfrak{L}=\sqrt{-g}\bigg(-\frac{1}{\kappa^2}R +\frac{1}{\kappa^2}2\epsilon_0 (\nabla \phi)^2 + \frac{1}{\mu_0}e^{-2 \alpha_0 \phi } F^2 \bigg),
\end{equation}
where $\kappa^2=8\pi G/c^4$, $c$ represents the speed of light, $G$ denotes the gravitational constant, and $\mu_0$ is the vacuum permeability. The scalar field is denoted by $\phi$; $R$ is the Ricci invariant and $g$ is the determinant of the metric. The arbitrary parameter $\alpha_0$ defines the theory, for example, $\alpha_0^2=3,1,0$ for Kaluza-Klein, Superstrings or Einstein-Maxwell, and $\epsilon_0=\pm 1$ if the scalar field is successively dilatonic or phantom-like.\footnote{In this article, we will focus exclusively on the examination of dilatonic scalar fields. This is due to the fact that, in the case of phantom scalar fields, in the equatorial plane, the throat and the ring singularity are identical.}
The corresponding field equations (FE) of the Lagrangian (\ref{LagrangianoTesisUnidades}) are
\begin{subequations}\label{EcuacionesDeCampoOriginales}
\begin{align}
    &\nabla_\mu \left( e^{-2\alpha_0 \phi} F^{\mu \nu} \right)=0 \label{Eq:Campo1}, \\
    &\epsilon_0 \nabla^2 \phi+\frac{\alpha_0}{2} \sigma_0 \left( e^{-2\alpha_0 \phi} F^{2} \right)=0 \label{Eq:Campo2},\\
    &R_{\mu \nu}=2\epsilon_0 \nabla_\mu \phi \nabla_\nu \phi \notag \\ 
    & \qquad + 2 \sigma_0 e^{-2\alpha_0 \phi} \left( F_{\mu \sigma} \tensor{F}{_\nu}{^\sigma} -\frac{1}{4} g_{\mu \nu } F^2 \right) \label{Eq:Campo3},
\end{align}
\end{subequations}
where $\nabla^2= \nabla^\mu \nabla_\mu$, and $\sigma_0 \equiv \frac{8\pi G}{\mu_0 c^4}$.

%\subsection{Coordinates and relations of the parameters}
%{\it The solution.}
The space-time under consideration is characterized by the presence of two Killing vectors, namely $\partial_t$ and $\partial_\varphi$, which correspond to the properties of stationarity and axial symmetry, respectively. The Weyl anzat metric expressed in spheroidal Oblates($+$)/Prolates($-$) coordinates $(x, y)$ is defined as

{\setlength{\abovedisplayskip}{-10pt}
 \setlength{\abovedisplayshortskip}{-10pt}
\begin{multline}\label{ds sp}
    ds^2 = -f\left( cdt-\omega d \varphi \right)^2
     + \frac{(L_{\pm})^2}{f} \bigg( (x^2\pm1)(1-y^2) d\varphi^2 \\
     +(x^2\pm y^2) e^{2k} \left\{ \frac{dx^2}{x^2\pm 1} +\frac{dy^2}{1-y^2} \right\} \bigg).
\end{multline}
}
where the metric functions $\{f, \omega, \kappa\}$ depend on the Weyl-coordinates $\left(\rho=(L_{\pm})\sqrt{(x^2\pm1)(1-y^2)}, z=(L_{\pm})xy\, \right)$, $\rho \in [0,\infty)$, $\{ z ,x \}\in \mathbb{R}$, $y \in [-1,1]$, and $(L_{\pm}) \geq 0$. 
We are examining two scenarios: the fisrt scenario pertains to the sub-extreme (S-E:$-$) condition
\begin{align}\label{Sub-Extreme}
|\mathcal{M}_\infty|^{2}&>a^2+Q_{L}^{2}+H_{L}^{2} ,\\
|\mathcal{M}_\infty|^{2}&=L_{-}^2+a^2+Q_{L}^{2}+H_{L}^{2},
\end{align}
the second one corresponds to the super-extreme (SU-E:$+$) case
\begin{align}\label{Super-Extreme}
|\mathcal{M}_\infty|^{2}&<a^2+Q_{L}^{2}+H_{L}^{2} ,\\
L_{+}^2+|\mathcal{M}_\infty|^{2}&=a^2+Q_{L}^{2}+H_{L}^{2}.
\end{align}
where \(a = J_\infty / |\mathcal{M}_\infty|\) represents the angular momentum per unit effective mass, with \(\mathcal{M}_\infty = l_1 + i\,N_\infty\), \(l_1\) a parameter with dimensions of length, and \(Q_L = Q_\infty\), \(H_L = H_\infty\) the electric and magnetic geometric charges, respectively, while \(N_\infty\) is the NUT parameter. The quantities \(Q_\infty, H_\infty, N_\infty, J_\infty\) are the conserved invariant charges defined in the classical sense as in \cite{Komar:1958wp,Nedkova:2011hx,Clement:2015aka,Clement:2022pjr,BallonBordo:2019vrn,Misner:1963fr,Manko:2005nm}. In general, the Komar mass \(M_\infty\) coincides with \(l_1\); in our particular family of solutions, however, \(l_1\) serves solely as a parameter of the wormhole throat.
Finally, the Boyer-Lindquist coordinates $(r,\theta)$ are related to the previous coordinates as
{\setlength{\abovedisplayskip}{0pt}
 \setlength{\abovedisplayshortskip}{0pt}
\begin{equation}\label{BoyerLindquistSpheroidalCordiantes}
    (L_{\pm}) x=r-l_1, \qquad y=\cos{\theta},
\end{equation}
}
where $r \in (-\infty,-l_1]\cup[l_1,\infty)$, $\theta\in [0,\pi]$, and the variable $l_1=r_s/2$, where $r_s$ represents the Schwarzschild radius.

In the first section, we will construct the WCCC (explained in \cite{axioms14110831}) from the topological point of view, and we will project this representation onto our spheroidal coordinates to apply to an exact solution of (\ref{EcuacionesDeCampoOriginales}).

In section two, we examine the second class of solutions discussed in \cite{Bixano:2025bio,Bixano:2025jwm,DelAguila:2015isj} and show that the S-E configuration corresponds to a black hole. For the region outside the black hole event horizon, we take $x>1$ or $r>l_1+(L_{-})$, whereas the interior region is defined by $x\in [-l_1/L_{-},1)$ or $r\in [0,l_1+L_{-})$.

In the SU-E case, this setup represents a wormhole. For the region $|y| > y_1$, the throat is located at $x_G = 0$, so we identify $x>0$ or $r>l_1$ with one universe, while $x<0$ or $r<-l_1$ may be interpreted either as a second universe or as a distant portion of the same universe. When $|y| \le y_1$, the throat shifts to a position with $x_G \neq 0$.

%------------------------------------------------------------
%%%%%%%%%%%%%%%%%%%%%%%%%%%%%%%%%%%%%%%%%%%%%%%%%%%%%%%%%%%%%%%%%%%%%%%%%%%%%%%%%%%%%%%%%%%%%%%%%%%
%%%%%%%%%%%%%%%%%%%%%%%%%%%%%%%%%%%%%%%%%%%%%%%%%%%%%%%%%%%%%%%%%%%%%%%%%%%%%%%%%%%%%%%%%%%%%%%%%%%
\section{Wormhole Cosmic Censorship Conjeture}
%------------------------------------------------------------
%------------------------------------------------------------
The goal of this section is to turn the informal idea that \emph{the throat encloses the defects} into a precise and reusable formalism. The central observation is that, in stationary axisymmetric spacetimes, several types of physically relevant pathologies, such as curvature singularities, CTCs, and even Killing horizons, can be collected into a single \emph{defect set}. The wormhole throat, defined intrinsically as a strict minimum of the areal functional, then serves as a canonical separating surface. We introduce a concrete framework that makes the notion of enclosing defects mathematically precise, provides sharp, verifiable conditions for explicit solutions, and fits naturally with the slice-by-slice Penrose compactifications employed later in the paper.
%______________________________________________________________________________________________________
\subsection{Defect set: curvature, CTCs, and horizons}
\label{subsec:defect_set}
%______________________________________________________________________________________________________
We consider a stationary, axisymmetric spacetime $(M,g)$ admitting the Killing vector fields $\partial_t$ and $\partial_\varphi$, and introduce spheroidal coordinates $(x,y)$ with $y\in[-1,1]$ and $\varphi\sim\varphi+2\pi$. Let $\Sigma$ be a hypersurface of constant $t$ lying in the regular region.
\paragraph{Curvature defects:}
We shall introduce a curvature measure given by some spacetime scalar invariant $K$. A point $p\in M$ in the spacetime manifold $M$ is a curvature defect if $K(p)=\pm \infty$. 
%......................
\paragraph{Chronology defects (CTCs).}
In an axisymmetric spacetime with a Killing vector $\partial_\varphi$, closed $\varphi$-orbits arise as the natural candidates for closed curves. Define the axial norm
\begin{equation}\label{eq:chi_def}
\chi=g(\partial_\varphi,\partial_\varphi)=g_{\varphi\varphi}.
\end{equation}
If $\chi<0$ at some point, then the $\varphi$-orbits are timelike, implying that the geometry contains closed timelike curves passing through that point. Consequently, the chronology-violating set is
\begin{equation}\label{eq:CTC_set}
\mathcal V_{\rm CTC}=\{p\in M:\ \chi(p)<0\}.
\end{equation}
%......................
\paragraph{Horizon defects.}
Even though our SU-E wormhole sector does not contain an event horizon, it is still conceptually helpful to treat possible horizons within the same defect classification framework. In stationary spacetimes, a natural quasi-local candidate is a (Killing) horizon tied to the stationary Killing vector field, or, more broadly, a horizon generated by
\begin{equation}\label{eq:Killing_combo}
K_{t\varphi}=\partial_t+\Omega\,\partial_\varphi,
\end{equation}
for some constant $\Omega$ at the surface $\mathscr{H}$. A Killing horizon is defined as a null hypersurface $\mathscr{H}$ for which $g(K_{t\varphi},K_{t\varphi})=0$ and the vector field $K_{t\varphi}$ is everywhere normal to $\mathscr{H}$. We gather all such potential horizons into the set $\mathcal V_{\rm H}$.
%......................
\paragraph{Unified defect set and regular domain.}
We define the \emph{defect set} as
\begin{equation}\label{eq:defect_set_union}
\mathcal V=\{p:\ K(p)=\pm \infty\}\ \cup\ \mathcal V_{\rm CTC}\ \cup\ \mathcal V_{\rm H},
\end{equation}
and the domain as
\begin{equation}\label{eq:regular_domain}
\mathcal D=M\setminus \mathcal V.
\end{equation}
By construction, $\mathcal V\cap\mathcal D=\emptyset$.
%______________________________________________________________________________________________________
\subsection{Areal throat as a canonical separating surface}
\label{subsec:areal_throat_formal}
%______________________________________________________________________________________________________
On $\Sigma$, consider the closed two-dimensional surfaces $\mathcal S_x=\{t=\mathrm{const},\,x=\mathrm{const}\}$, which are parametrized by $(y,\varphi)$. Their associated area functional is
\begin{equation}\label{eq:Areal_formal}
\mathcal A(x)=\int_{0}^{2\pi} d\varphi  \int_{-1}^{1}\!dy\,\sqrt{g_{yy}(x,y)\,g_{\varphi\varphi}(x,y)}.
\end{equation}
We characterize the wormhole throat as a \emph{strict} local minimum:
\begin{equation}\label{eq:areal_min_formal}
\left.\frac{d\mathcal A}{dx}\right|_{x=x_G}=0,
\qquad
\left.\frac{d^2\mathcal A}{dx^2}\right|_{x=x_G}>0,
\end{equation}
and refer to the associated minimal surface as
\begin{equation}\label{eq:throat_surface}
\mathcal T=\mathcal S_{x_G}.
\end{equation}
This definition is strictly geometric and does not assume any specific deformation. The position of the throat, $x_G$, is fully fixed by the precise metric functions entering \eqref{eq:Areal_formal}.
%______________________________________________________________________________________________________
\subsection{\texorpdfstring{Meridional censorship radii: compressing all pathologies into $x_\ast(y)$}{Meridional censorship radii: compressing all pathologies into x_*(y)}}
\label{subsec:xstar}
%______________________________________________________________________________________________________
Because the spacetime is axisymmetric, we can analyze the geometry on a meridional slice by keeping $\varphi$ constant, in other words, we will work with the projection of the defect set onto a meridional plane. This is equivalent to working on the two-dimensional quotient by the axial Killing orbits. Fix an arbitrary $y\in[-1,1]$ and examine the ray $\{(x,y): x>0\}$. We then define the \emph{censorship radius} $x_\ast(y)$ as the smallest value of $x$ beyond which the geometry along that meridional direction becomes smooth and free of any violations of chronology:
\begin{equation}\label{eq:xstar_def}
x_\ast(y)=\inf\Big\{x>0:\ (x',y)\in\mathcal D\ \text{for all}\ x'\ge x\Big\}.
\end{equation}
Thus, the inner region $0 \le x < x_\ast(y)$ coincides precisely with the portion of the meridional ray that lies outside the regular domain. It encompasses curvature singularities and/or violations of chronology and/or possible horizons.

Let $x_s(y)$ and $x_v(y)$ denote, respectively, the limiting curves associated with curvature and with chronology (for instance, the smallest value of $x$ beyond which $K$ is finite, and the smallest value of $x$ beyond which $\chi\ge0$). When relevant, let $x_h(y)$ denote the smallest $x$ beyond which $g(K,K)\neq0$ for potential horizons. By definition, we then have
\begin{equation}\label{eq:xstar_max}
x_\ast(y)=\max\{x_s(y),\,x_v(y),\,x_h(y)\}.
\end{equation}
Consequently, this single function $x_\ast(y)$ consolidates all the relevant defect thresholds into a single entity.
%______________________________________________________________________________________________________
\subsection{Enclosure as a topological separation statement}
\label{subsec:topological_enclosure}
%______________________________________________________________________________________________________
We now give a precise definition of what it means for the throat to \emph{enclose} the defects.
\begin{definition}[Topological enclosure by the throat]
\label{def:enclosure}
Let $\mathcal T\subset\Sigma$ be an embedded, closed two-dimensional surface (topologically equivalent to $S^2$).We say that the throat \emph{encloses} the defect set on $\Sigma$ if
\begin{equation}\label{eq:enclosure_set}
(\Sigma\cap\mathcal V)\subset \mathrm{Int}_\Sigma(\mathcal T),
\end{equation}
where $\mathrm{Int}_\Sigma(\mathcal T)$ denotes the interior component of $\Sigma\setminus\mathcal T$.
\end{definition}
The division into an interior and an exterior corresponds to the usual Jordan–Brouwer separation property for embedded $S^2$ surfaces: the set $\Sigma\setminus\mathcal T$ decomposes into two connected components, an exterior that includes the asymptotic region(s) and an interior whose boundary is $\mathcal T$.
\begin{proposition}[Operational enclosure criterion in meridional variables ($x,y$)]
\label{prop:operational_enclosure}
Suppose that for every $y\in[-1,1]$, the regular region along the meridional ray is given by $x\ge x_\ast(y)$, and that the throat crosses this ray at the point $x = x_G(y)$. If
\begin{equation}\label{eq:enclosure_ineq}
x_\ast(y)<x_G(y)\qquad \text{for all}\qquad y\in[-1,1],
\end{equation}
then the throat serves as an enclosure for all defects on $\Sigma$ in the sense of definition~\ref{def:enclosure}.
\end{proposition}
\begin{proof}
Fix an arbitrary $y$. By the definition of $x_\ast(y)$, every point with $0\le x<x_\ast(y)$ lies outside the regular domain, in other words, all such points belong to the excluded set arising from curvature, chronology, or horizon defects. The inequality \eqref{eq:enclosure_ineq} further ensures that the interval $0\le x<x_\ast(y)$ is strictly on the interior side of the throat intersection $x=x_G(y)$. Since this is true for all $y$, the entire meridional projection of the defect set onto $\Sigma$ is contained in the region enclosed by $\mathcal T$. Consequently, we have $(\Sigma\cap\mathcal V)\subset \mathrm{Int}_\Sigma(\mathcal T)$.
\end{proof}
\begin{remark}[Horizon inclusion]
\label{rem:horizon_inclusion}
The presence of $\mathcal V_{\rm H}$ in \eqref{eq:defect_set_union} leads to the following operational interpretation: whenever candidate horizons are present, the same inequality \eqref{eq:enclosure_ineq} constrains them to be located strictly inside the throat. In the specific wormhole considered here one finds $\mathcal V_{\rm H}=\emptyset$, but the formalism is constructed to remain valid for related families in which horizons can occur.
\end{remark}
%______________________________________________________________________________________________________
\subsection{Slice-wise causal compactification}
\label{subsec:slicewise_penrose_formal}
%______________________________________________________________________________________________________
For the Penrose compactification, one considers the two-dimensional Lorentzian subsector defined by $\varphi=\varphi_0$, $y=y_0=\mathrm{const}$, and expresses the induced metric in the $(t,x)$ coordinates as
\begin{equation}\label{eq:tx_sector_formal}
ds^2=-f(x,y_0)dt^2+G(x,y_0)\,dx^2,
\quad x\in[x_\ast(y_0),+\infty),
\end{equation}
with $G(x,y_0)>0$ and $f(x,y_0)>0$ along the regular curve.
Let the tortoise coordinate $l={l\Big|}_{y_0}(x)$ by requiring
\begin{equation}\label{eq:tortoise_formal}
ds^2=-f(x,y_0)dt^2+f(x,y_0)\,dl^2,
\end{equation}
therefore
\begin{equation}\label{eq:dl_dx_general_formal}
\frac{dl_{y_0}}{dx}=\sqrt{\frac{G(x,y_0)}{f(x,y_0)}}>0
\qquad \text{on }[x_\ast(y_0),+\infty).
\end{equation}
Therefore, $l_{y_0}(x)$ is strictly increasing along the regular branch. Consequently, for any causal curve restricted to the regular two-dimensional sector, moving toward smaller values of $x$ corresponds monotonically to moving toward smaller values of $l$. Because the regular manifold starts at $x = x_\ast(y_0)$, every such causal curve must reach the boundary $x = x_\ast(y_0)$ and, in particular, the throat at $x = x_G(y_0) \ge x_\ast(y_0)$, before it can enter the excluded interior that contains the defect set. \emph{This is the precise sense in which one encounters the throat first and proceeds to the other asymptotic region, while the defect set remains unreachable from within the regular wormhole geometry.}
%========================================================
%========================================================
\subsection{WCCC as a topological--causal principle}
\label{subsec:general_WCCC_topological}

We are now in a position to propose a wormhole counterpart of cosmic censorship that is formulated intrinsically on $\Sigma$, expressed in terms of meridional variables in an operational way, and consistent with Penrose compactification performed slice by slice.

\begin{conjecture}[Wormhole Cosmic Censorship Conjecture, topological form]
\label{conj:WCCC_topological}
Let $(\mathcal M,g)$ denote a stationary, axisymmetric spacetime, and let $\Sigma$ be a spacelike hypersurface equipped with meridional coordinates $(x,y)$ and a defect set $\mathcal V\subset\Sigma$ as specified in \eqref{eq:defect_set_union}. Assume that the regular region $\mathcal D=\Sigma\setminus\mathcal V$ admits a censor radius $x_\ast(y)$ defined by \eqref{eq:xstar_def}. Suppose further that there exists an embedded spherical throat $\mathcal T\subset\Sigma$ whose meridional projection is given by a graph $x=x_G(y)$, characterized intrinsically as a strict minimizer of the areal functional on $\Sigma$, satisfying $\delta\,\mathrm{Areal}=0$ and $\delta^2\mathrm{Areal}>0$ within the admissible class. Then, for any physically admissible wormhole configuration,
\begin{equation}\label{eq:WCCC_inequality}
x_\ast(y)<x_G(y)\qquad \text{for all}\qquad y\in[-1,1],
\end{equation}
so that the throat encloses the entire defect set on $\Sigma$:
\begin{equation}\label{eq:WCCC_enclosure}
(\Sigma\cap\mathcal V)\subset \mathrm{Int}_\Sigma(\mathcal T).
\end{equation}
\end{conjecture}
\begin{remark}[Causal reading]
\label{rem:WCCC_causal_reading}
In the presence of the slice-wise tortoise coordinate \eqref{eq:dl_dx_general_formal}, the inequality \eqref{eq:WCCC_inequality} leads to the following operational interpretation: within each sector of fixed $y_0$, any causal curve originating in the regular exterior necessarily meets the throat (located at $x=x_G(y_0)$) \emph{before} it can approach the excised interior region $x<x_\ast(y_0)$. In other words, the defect set cannot be accessed from the regular wormhole geometry, and the compactified Penrose diagram accordingly features a finite central strip delimited by the throat lines.
\end{remark}
\begin{remark}[What WCCC does not claim]
\label{rem:WCCC_not_claim}
The conjecture does not claim that defects are absent. Instead, it states that whenever defects (such as curvature blow-ups, chronology violations, or horizons) appear in a stationary wormhole configuration, they are \emph{topologically sequestered} within the throat. That is, they are isolated from the smooth wormhole exterior by an embedded $S^2$ that minimizes area. This serves as the wormhole counterpart of the principle that singularities are concealed behind horizons in the black-hole formulation of cosmic censorship.
\end{remark}
%%%%%%%%%%%%%%%%%%%%%%%%%%%%%%%%%%%%%%%%%%%%%%%%%%%%%%%%%%%%%%%%%%%%%%%%%%%%%%%%%%%%%%%%%%%%%%%%%%%
%%%%%%%%%%%%%%%%%%%%%%%%%%%%%%%%%%%%%%%%%%%%%%%%%%%%%%%%%%%%%%%%%%%%%%%%%%%%%%%%%%%%%%%%%%%%%%%%%%%
\section{Solution under investigation}
%------------------------------------------------------------
%------------------------------------------------------------
The solution we will examine corresponds to the metric functions determined by $\lambda_6=\lambda_0 \frac{y}{(x^2\pm y^2)}$, which constitute a solution to the FE (\ref{EcuacionesDeCampoOriginales})
{\small
\begin{subequations}\label{SolucionLambdaCombinada}
\begin{align}
        \phi&=-\frac{\lambda_6}{\alpha_0}
        , \label{CampoEscalar Lambda 6}\\
        f&=f_0=1,\\
        \omega_6 &=\frac{ L_\pm \lambda_0 }{f_0} \frac{x(1-y^{2})}{x^{2}\pm y^{2}}, \label{Omega Lambda6}\\
        k_{\lambda_6} &= -k_{0} \lambda_0 ^2  \frac{(1-y^2)}{4(x^2\pm y^2)^4} \bigg(  8x^2y^2(x^2 \pm 1) \notag \\
        &-(x^2 \pm y^2)^2(1-y^2)  \bigg) \label{k Lambda6} \\
        A_{\varphi6} &=-\frac{\sqrt{f_0}}{2 \kappa_0 \sqrt{\sigma_0}} \frac{\omega_6}{L_\pm} e^{-\lambda_{6}}, \label{A3 Lambda6} \\
        A_{t6}&=\frac{\sqrt{f_0}}{2\kappa_0 \sqrt{\sigma_0}} (e^{-\lambda_{6}}-1),
\end{align}
\end{subequations}
}
where, as before, the upper sign corresponds to the super-extreme case (SU-E), and the lower sign corresponds to the sub-extreme case (S-E), $\{\lambda_0,k_0,\kappa_0,f_0\}$ as integration constants, we need to satisfy the parameter constraint\footnote{For a detailed derivation of the calculations, see \cite{axioms14110831}} $\alpha_0^2(4k_0+1)-4\epsilon_0=0$ to fulfil \eqref{Eq:Campo3}. An important point for this solution is that the conserved charge invariants are given by $Q_\infty = 0 = H_\infty$, $M_\infty = 0$, $J_\infty = -\frac{\lambda_0 \, L_{\pm}^2}{2f_0}$, and the NUT parameter vanishes, $N_\infty = 0$, as reported in \cite{Bixano:2025bio}. 
%______________________________________________________________________________________________________
%\subsection{The behaviour}
%______________________________________________________________________________________________________
%{\it The behaviour}. 
The works \cite{DelAguila:2015isj,Bixano:2025jwm,Bixano:2025bio} illustrate that these types of solutions exhibit a ring singularity at $x=y=0$ ($\rho=L,z=0$) or ($r=l_1 , \theta=\pi/2$) under SU-E considerations. However, for the S-E considerations, the Ricci and Kretschmann scalars are proportional to $1/(x^2 \pm y^2)^\iota$, where $\iota$ is 4 and 12, respectively. Consequently, the same ring singularity can be identified at $x=0\quad \& \quad y=0$, and two superficial singularities emerge at $x=\pm y$ or $r=l_1\pm (L_{\pm})\cos{\theta}$. The solution being analysed is asymptotically flat, hence, as $x \rightarrow \infty$ in sub-extreme scenario, or $x \rightarrow \pm \infty$  in super-extreme scenario, the set $\{\phi,\omega,k,A_t,A_\varphi\}$ converges towards zero and $\rho \approx r^2$.
%
%______________________________________________________________________________________________________
%\subsection{Ring singularity}
%______________________________________________________________________________________________________
Let us focus on the hypersurface defined by $\varphi=\varphi_0$ (so that $d\varphi=0$) and assume the SU-E case, under these conditions, \eqref{ds sp} becomes:
{\small
\begin{equation}\label{eq:ds_reduced_xy}
ds^2=-fdt^2+\frac{L_+^2}{f}\,(x^2+y^2)\,e^{2k}
\left(\frac{dx^2}{x^2+1}+\frac{dy^2}{1-y^2}\right).
\end{equation}
}
\emph{Throughout the computation we will assume $k_0>0$, meaning that our scalar field is of a dilaton-like type.}
The meridional point \((x,y) = (0,0)\), obtained by projecting onto the hypersurface of constant \(\varphi\), represents the ring defect in the complete spacetime. Specifically, on the equatorial section \(y = 0\),
\begin{equation}\label{eq:k_equator_again}
k(x,0)=\frac{k_0\lambda_0^2}{4x^4}\xrightarrow[x\to0]{}+\infty.
\end{equation}
%.....................
\begin{lemma}[Exponential domination]
\label{lem:superexp}
Let $c>0$ and $m>0$. Then
\begin{equation}
\label{eq:superexp}
\lim_{x\to 0^+} x^m e^{c/x^4}=+\infty.
\end{equation}
\end{lemma}
%.....................
\begin{proof}
We compute
\[
\ln\!\big(x^m e^{c/x^4}\big)=m\ln x + \frac{c}{x^4}.
\]
As $x \to 0^+$, we observe that $\ln x \to -\infty$, whereas $\frac{c}{x^4} \to +\infty$ and dominates any logarithmic divergence in growth rate. Therefore, the right-hand side diverges to $+\infty$, which yields \eqref{eq:superexp}.
\end{proof}
%.....................
Along $y=0$, the spatial $xx$ component of \eqref{eq:ds_reduced_xy} is given by
\begin{equation}
\label{eq:gxx_equator}
g_{xx}(x,0)=\frac{L_+^2}{f_0}\,\frac{x^2}{x^2+1} e^{\left(\frac{k_0\lambda_0^2}{2x^4}\right)}.
\end{equation}
Using Lemma~\ref{lem:superexp} with the choices $m=2$ and $c=k_0\lambda_0^2/2>0$, we obtain
\begin{equation}
\label{eq:gxx_blowup}
\lim_{x\to0^+} g_{xx}(x,0)=+\infty.
\end{equation}
Consequently, the equatorial slice cannot be smoothly continued as a regular Lorentzian manifold across $x=0$,  the singular set in the meridional plane is
\begin{equation}\label{eq:xs_def}
x_s=0 \qquad (y=0).
\end{equation}
%______________________________________________________________________________________________________
%\subsection{Event horizons}
%______________________________________________________________________________________________________
%{\it Event horizons.}
To determine the event horizons, we use the Killing vector $K_t=\partial_t + \Omega\partial_\psi$, with $\Omega$ being constant across the horizon $\mathscr{H}$. This is accomplished by calculating the norm of the vector $K_t$ and setting it equal to zero
\begin{align*}
    &g(K_t,K_t)=g_{tt}+2\Omega g_{t\psi}+\Omega^2g_{\psi \psi}=0, \\
    & \text{implies}\, \Omega_{\pm}=-\frac{g_{t\psi}}{g_{\psi \psi}}\pm \frac{\sqrt{-(g_{\psi \psi}g_{tt}-(g_{t\psi})^2)}}{g_{\psi \psi}}.
\end{align*}
Imposing 
\[
(g_{t\psi})^2-g_{\psi \psi}g_{tt} =\rho=-L_{\pm}\sqrt{(x^2 \pm 1)(1-y^2)}=0,
\]
where $f=1$, we obtain a non degenerate 
\[
\Omega=-g_{t\psi}/g_{\psi \psi}=(\omega|_{\mathscr{H}})^{-1}.
\]
\paragraph{Super-extreme case:}
In the SU-E scenario, \textit{no horizon exists}, because we see that $\rho=0 \Leftrightarrow y=\pm 1$ and substituting into \eqref{Omega Lambda6}, we find that $\omega|_{\mathscr{H}}=\omega(x,y=\pm 1)=0$,then the constant $\Omega$ is not well-defined on $\mathscr{H}$.
\paragraph{Sub-extreme case:}
On the other hand, in the S-E \textit{the event horizon is in} $x=\pm 1 \, \,\text{implies} \, \, r=l_1\pm L_{-}$ with $\omega|_{\mathscr{H}}=\omega(x=\pm 1,y)=L_{-} \lambda_0:\text{constant}$, then the \textit{sub-extreme} case is a Black Hole with horizon and $\Omega$ given by
\begin{equation}\label{eq:EvenHorizon-SE}
    x_h=\pm 1, \qquad \Omega \Big|_{\mathscr{H}}=\frac{1}{\lambda_0\, L_{-}}.
\end{equation}
The surface gravity at the event horizon is determined by calculating 
\begin{equation}\label{eq:SurfaceHGravity-SE}
    \varsigma=\lim_{p\rightarrow\mathscr{H}}\sqrt{\gamma ^{ij}(\partial_{i} \Xi ) (\partial_{j} \Xi )}=e^{-k_0 \lambda_0^2/4}/(L_{-} \lambda_0),
\end{equation}
where $\gamma^{ij}$ is the metric restricted to the ($x,y$)-subspace and $\{i,j\}=x,y$. 

\begin{figure}[h!]
  \centering
  \includegraphics[
  width=\linewidth,
  height=\textheight,
  keepaspectratio
]{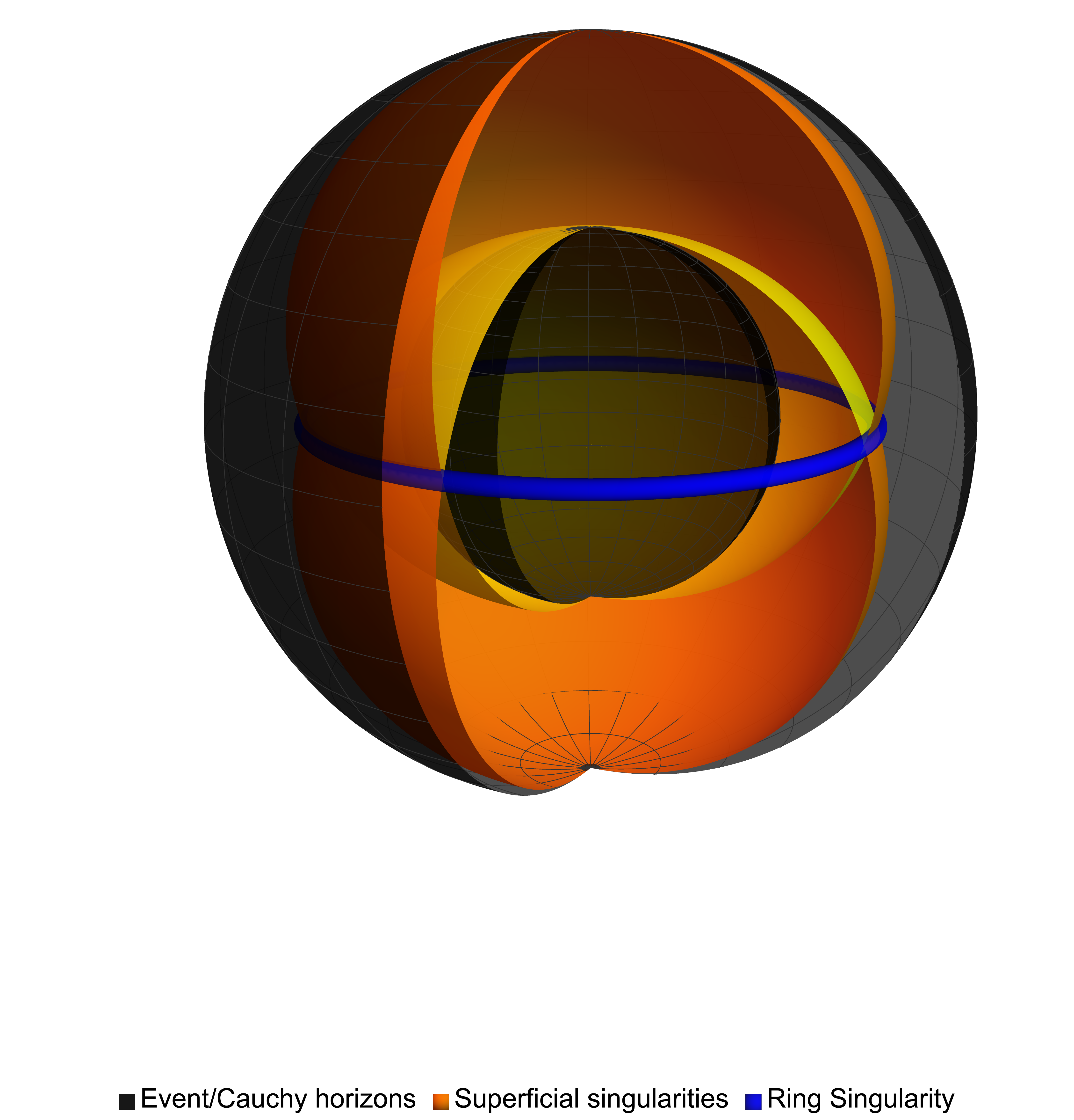}
\caption{The black outer layer signifies the event horizon, while the inner black layer denotes the Cauchy surface, the blue torus symbolizes the ring singularity, while the yellow layer indicates a surface singularity and the orange signifies another one. This figure is a schematically Cosmic Censorship illustration.}
    \label{fig:BHStructure}
\end{figure}

The schematic depiction of this black hole can be found in Figure \ref{fig:BHStructure}.
%
%______________________________________________________________________________________________________
%\subsection{Closed Time-like Curves}
%______________________________________________________________________________________________________
%{\it Closed Time-like Curves}
In order to identify the CTCs, it is important to recognize that within the coordinate $\varphi \thicksim \varphi+n\pi$, there exists a closed curve. Therefore, the variable $\varphi$ acts as a time function when $g(\partial_\varphi,\partial_\varphi)=g_{\varphi \varphi}=(\rho^2-f^2\omega^2)/f<0$. In other words, we need to analize
\begin{equation}\label{eq:gphiphi_omega6}
g_{\varphi\varphi}(x,y)=\frac{L_{\pm}^2}{f_0}(1-y^2)\left[
(x^2\pm 1)-\lambda_0^2\,\frac{x^2(1-y^2)}{(x^2 \pm y^2)^2}
\right],
\end{equation}
where $L_{\pm}>0$, $f_0>0$, $\lambda_0>0$, and $|y|\leq 1$. 
%
%**********************************************
\subsubsection{S-E case}
Let us start with the S-E (Lower sign), we will examine it step by step.
%**********************************************

For $y=\pm1$, axial degeneration implies $g_{\varphi\varphi}(x,\pm1)=0$. Consequently, to diagnose the presence of CTCs we must restrict to $|y|<1$, where the azimuthal trajectories have nonzero length.

For $|x|<1$ we have $x^2-1<0$. Moreover, for $|y|<1$, the second term in \eqref{eq:gphiphi_omega6}, namely $\lambda_0^2\,\frac{x^2(1-y^2)}{(x^2-y^2)^2}$, is non-negative. Hence, $g_{\varphi\varphi}(x,y)<0$.

Now, evaluating \eqref{eq:gphiphi_omega6} at $x=\pm1$, we obtain $g_{\varphi\varphi}(\pm1,y)=-\frac{L_-^2}{f_0}\lambda_0^2$, therefore $g_{\varphi\varphi}(x,y)<0$ at the event horizon for all $\lambda_0,L_{-}\in \mathbb{R}$ and $|y| < 1$.

%--------------------------------------------------------
Finally, from \eqref{Omega Lambda6} we observe that the key term determining whether $g_{\varphi \varphi}$ is negative or positive is
\[
\chi_y(x)=(x^2-1)-\lambda_0^2\,\frac{x^2(1-y^2)}{(x^2-y^2)^2}
\]
Hence, we restrict to $x>1$. Since $|y|<1$, it follows that $x^2-y^2>0$, and consequently $\chi_y(x)$ defines a smooth—indeed, real-analytic—function on the interval $x\in(1,\infty)$.
\begin{proposition}[Existence of a real outer root $x_v(y)>1$]
\label{prop:exist_xv}
Assume $\lambda_0\neq 0$ and choose any $y\in(-1,1)$. Then there is at least one real number
\begin{equation}\label{eq:xv_exists_statement}
x_v(y)\in(1,\infty)
\qquad\text{such that}\qquad
\chi_y\big(x_v(y)\big)=0,
\end{equation}
equivalently $g_{\varphi\varphi}(x_v(y),y)=0$. In particular, the chronology-violating region
$g_{\varphi\varphi}(x,y)<0$ extends beyond $|x|=1$ for $|y|<1$ and has an
outer boundary located at some $|x|=x_v(y)>1$.
\end{proposition}
\begin{proof}
\textbf{Step 1 (negativity at $x=1$).}
We know that $\chi_y(\pm1)=-\frac{\lambda_0}{1-y^2}<0$ for $|y|<1$ and $\lambda_0 \neq0$. 
On the other hand, we can see that 
{\small
\begin{equation*} 
\chi_y(x)=(x^2-1)-\lambda_0^2\,O\!\left(\frac{1}{x^2}\right)
\longrightarrow +\infty
\qquad (x\to+\infty).
\end{equation*}
}
In particular, there exists some $X>1$ for which $\chi_y(X)>0$.  Because $\chi_y$ is continuous on the closed interval $[1,X]$, and since $\chi_y(\pm1)<0$ while $\chi_y(X)>0$, the function must change sign on $[1,X]$. Hence, by the Intermediate Value Theorem, there is at least one point $x_v(y)\in(1,X)$ such that $\chi_y(x_v(y))=0$, where $x_v$ denotes the chronology bound associated with \eqref{eq:CTC_set} in the S–E setup.
\end{proof}
\begin{remark}[Definition of $x_v(y)$ as the \emph{outermost} root]
\label{rem:outermost_root}
The preceding argument ensures the existence of a real root in $(1,\infty)$. To choose a canonical threshold, one may define $x_v(y)$ to be the \emph{outermost} such root,
\begin{equation}
\label{eq:xv_outermost}
x_v(y)=\sup\{x>1:\ \chi_y(x)=0\}.
\end{equation}
\end{remark}

Therefore, for $\lambda_0\neq 0$ the exterior spacetime exhibits chronology violation outside the event horizon (in particular, a $g_{\phi\phi}<0$ region with closed timelike curves), so the exterior is not globally hyperbolic.
%Therefore, this BH has causal violations outside the event horizon for $\lambda_0 \neq 0$, and the Cosmic Censorship Conjeture is not satisfied.
%--------------------------------------------------------
%**********************************************
\subsubsection{SU-E case}
%**********************************************
As in the S–E scenario with $y=\pm1$, axial degeneration implies $g_{\varphi\varphi}(x,\pm1)=0$. Thus, to test for CTCs we restrict to $|y|<1$.

We now start by analysing the corresponding $\chi_y$ in the SU-E case, i.e.
\[
\chi_y(x)=(x^2+1)-\lambda_0^2\,\frac{x^2(1-y^2)}{(x^2+y^2)^2}.
\]
For all real $x,y$ one has
\begin{equation*}
(x^2+y^2)^2\ge (2|xy|)^2=4x^2y^2.
\end{equation*}
Consequently, for any $y\neq 0$,
{\small
\begin{equation*}
\frac{x^2}{(x^2+y^2)^2}\le \frac{1}{4y^2}  \quad \text{implies} \quad \frac{x^2(1-y^2)}{(x^2+y^2)^2}\le \frac{(1-y^2)}{4y^2}
\end{equation*}
}
Therefore, using this facts, we obtain 
\begin{align}
\chi_y(x)
&=(x^2+1)-\lambda_0^2\,\frac{x^2(1-y^2)}{(x^2+y^2)^2}
\nonumber\\
&\ge x^2+(1-\lambda_0^2\,\frac{1-y^2}{4y^2}).
\label{eq:B_lower_x}
\end{align}
Therefore, if
\[
1-\lambda_0^2\,\frac{1-y^2}{4y^2}>0 \quad \text{implies} \quad g_{\varphi \varphi}(x,y)>0,
\]
then no CTCs exist in this region. This condition is equivalent to
\begin{equation}\label{eq:y frontera CTCs}
    |y|\ > \frac{|\lambda_0|}{\sqrt{\lambda_0^2+4}} \qquad \text{for all}\qquad  x\in\mathbb{R}.
\end{equation}
We now establish a complementary estimate that holds uniformly in $y$. To this end, we make use of the simple inequality
\begin{equation*}
(x^2+y^2)^2\;\ge\;x^4,
\end{equation*}
we obtain, for $x\neq 0$,
\begin{equation*}
\frac{x^2}{(x^2+y^2)^2}\;\le\;\frac{1}{x^2}.
\end{equation*}
Since we also have $1 - y^2 \le 1$, it follows that
\begin{equation*}
\frac{x^2(1-y^2)}{(x^2+y^2)^2}\;\le\;\frac{1}{x^2}.
\end{equation*}
Substituting into $\chi_y(x)$ gives
\begin{equation*}
\chi_y(x)\;\ge\;(x^2+1)-\frac{\lambda_0^2}{x^2}.
\end{equation*}
If $|x|\ge|\lambda_0|$, then $\lambda_0^2/x^2\le 1$, and hence
\begin{equation*}
\chi_y(x)\;\ge\;(x^2+1)-1=x^2>0.
\end{equation*}
Hence, if
\begin{equation}\label{eq:x frontera CTCs}
|x|\ > |\lambda_0|,
\end{equation}
then $g_{\varphi\varphi}(x,y)>0\qquad \text{for all}\qquad  y\in(-1,1)$.

\begin{figure}[h!]
  \centering
  \includegraphics[
  width=\linewidth,
  height=\textheight,
  keepaspectratio
]{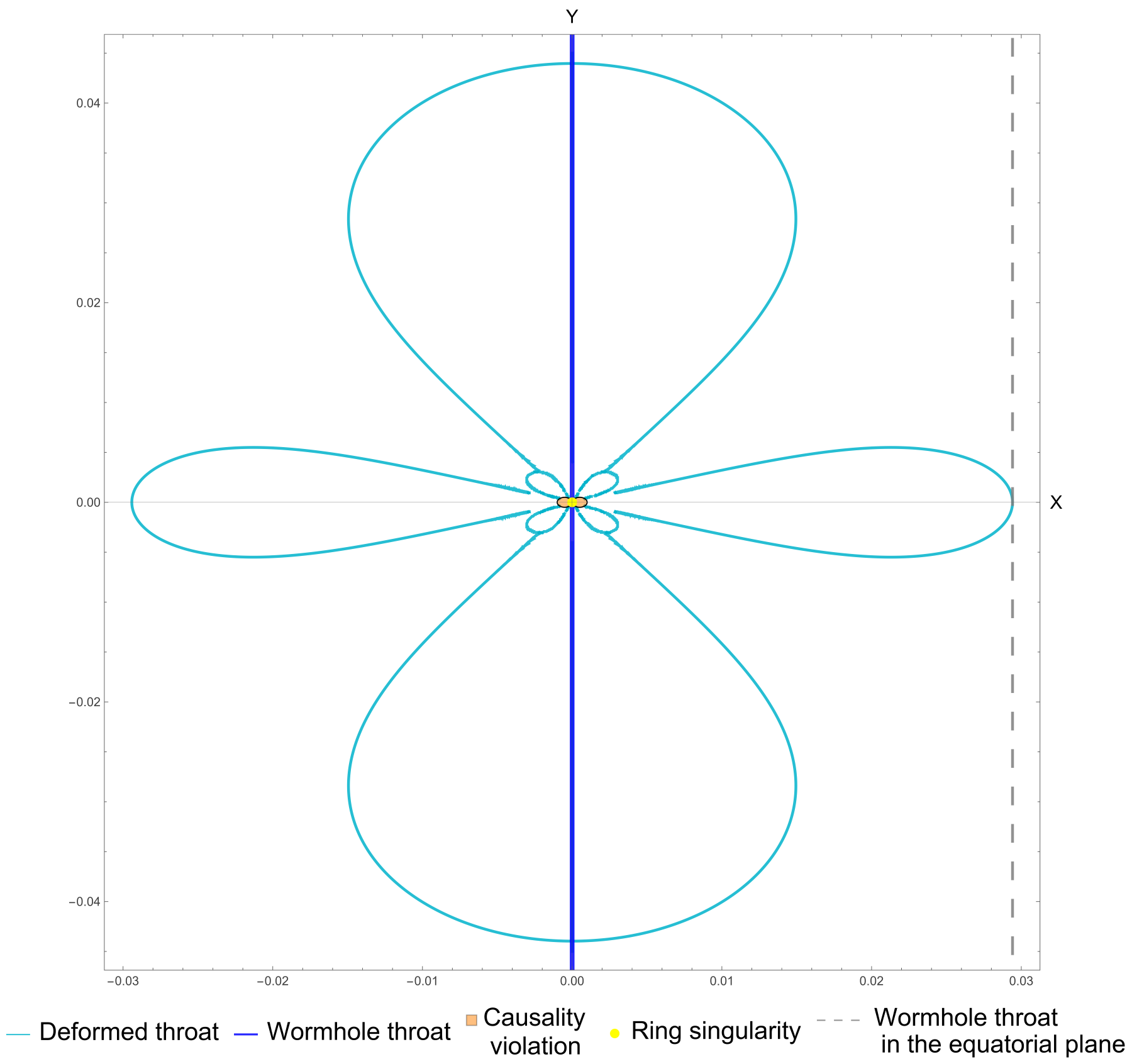}
\caption{This figure illustrates the numerical schematic of the WCCC employing the parameters $\lambda_0=10^{-3}$, $(L_{+}=1)$, and $k_0=3/4$ (in the context of Superstrings theory with a dilatonic field). The royal blue line denotes the throat ($x=0$) under the condition $y \notin [-45\lambda_0, 45\lambda_0]$. The outermost sky blue line signifies the deformed throat within the region $y \in [-45\lambda_0, 45\lambda_0]$, characterized by the conditions $\frac{d \, \text{Areal}}{d r}\big|_{x_G}=0$ and $\frac{d^2 \, \text{Areal}}{d x^2}\big|_{x_G}>0$. The dashed gray line depicts the throat in the equatorial plane ($y_0=0$), where it is found that $x_G \approx 0.02941> x_v \approx \lambda_0$. Furthermore, the yellow point signifies the ring singularity, while the two black circles represent the boundaries where Closed Timelike Curves (CTCs) emerge, indicating that CTCs are present within these two circles.}
\label{fig:ExpliGarganta}
\end{figure}

We have employed numerical methods to confirm these assertions, finding that causality violations occur in the region characterized by $|x_v| \approx \lambda_0 \ll 1$ and $y \in [-\lambda_0/2,\lambda_0/2]$, which is consistent with our theoretical findings, since from \eqref{eq:y frontera CTCs} and \eqref{eq:x frontera CTCs} we observe that for $\lambda \ll 1$, the region without CTCs is determined by $|y| > |\lambda_0|/2$ and $|x| < |\lambda_0|$. A graphical representation is shown in Figure \ref{fig:ExpliGarganta}.

We now restrict \eqref{eq:gphiphi_omega6} to the equatorial plane $y=0$. Noting that $(1-y^2)=1$ and $(x^2+y^2)^2=x^4$ in this case, we obtain
\begin{equation}
\label{eq:gphiphi_equator}
g_{\varphi\varphi}(x,0)=\frac{L_+^2}{f_0}\left[(x^2+1)-\frac{\lambda_0^2}{x^2}\right]
=\frac{L_+^2}{f_0}\,\frac{x^4+x^2-\lambda_0^2}{x^2}.
\end{equation} 
Consequently, the equatorial requirement $g_{\varphi\varphi}(x,0)>0$ is equivalent to 
\begin{equation}\label{eq:quartic_condition}
x^4+x^2-\lambda_0^2>0.
\end{equation}
By setting $u = x^2 > 0$, we obtain the inequality $u^2 + u - \lambda_0^2 > 0$, whose only positive root is
\begin{equation}\label{eq:uv_root}
u_v=\frac{-1+\sqrt{1+4\lambda_0^2}}{2}.
\end{equation}
Therefore, the chronological boundary is
\begin{equation}\label{eq:xv_def}
x_v=\sqrt{u_v}
=\sqrt{\frac{-1+\sqrt{1+4\lambda_0^2}}{2}}
\;>x_s=0.
\end{equation}
For $0<x<x_v$, the component $g_{\varphi\varphi}(x,0)$ is negative, so the axial circles $\varphi\sim\varphi+2\pi$ are timelike, which results in a breakdown of chronology. At $x=x_v$, the induced two-metric becomes degenerate at the \emph{equator} because $g_{\varphi \varphi}(x_v ,0)=0$, implying $\mathcal{A}(x)=0$.
\paragraph{Small-$\lambda_0$ expansion.}
Since $\sqrt{1+4\lambda_0^2}=1+2\lambda_0^2-2\lambda_0^4+O(\lambda_0^6)$, we obtain
{\small
\begin{equation}
\label{eq:xv_small_lambda}
u_v=\lambda_0^2-\lambda_0^4+O(\lambda_0^6),
\qquad
x_v=\lambda_0\left(1-\frac{\lambda_0^2}{2}+O(\lambda_0^4)\right).
\end{equation}
}
In particular, $x_v$ behaves as $x_v \sim \lambda_0$ when $\lambda_0 \ll 1$, satisfying \eqref{eq:x frontera CTCs}.

\begin{corollary}[Admissible domain for regular areal cross-sections]
\label{cor:admissible_domain}
If $x > x_v$, then $g_{\varphi\varphi}(x,y)\ge 0$ for all $y\in[-1,1]$, and in fact $g_{\varphi\varphi}(x,y)>0$ for every $y\in(-1,1)$. In contrast, at $x=x_v$ one finds $g_{\varphi\varphi}(x_v,0)=0$, which means that the induced two-metric on $\mathcal S_{x_v}$ becomes degenerate away from the axis, and thus $\mathcal S_{x_v}$ cannot represent a regular throat cross-section, where $\mathcal S_x:=\{t=\mathrm{const},\ x=\mathrm{const}\}$, and $\mathcal{A}(x)$ given by \eqref{eq:Areal_formal}.
\end{corollary}
\begin{proposition}[Rigorous inequality chain $x_G>x_v>x_s=0$ on the equatorial plane]
\label{prop:xG_gt_xv_gt_xs}
Assume that $0<\lambda_0<1$ and $k_0>0$. Let $x_G$ denote the intrinsic throat position, characterized as the point where the areal functional \eqref{eq:Areal_formal} attains a strict minimum when restricted to \emph{regular} two-surfaces $\mathcal S_x:=\{t=\mathrm{const},\ x=\mathrm{const}\}$.
\begin{equation}
\label{eq:xG_gt_xv_gt_xs}
x_G>x_v>x_s=0,
\end{equation}
where $x_v$ is defined exactly as in \eqref{eq:xv_def}.
\end{proposition}
\begin{proof}
By Corollary~\ref{cor:admissible_domain}, the induced metric on $\mathcal S_x$ is Riemannian for every $x>x_v$, whereas it becomes degenerate at $x=x_v$ (on the equator). Consequently, if there exists a \emph{regular} throat cross-section $\mathcal S_{x_G}$, it must fulfill $x_G>x_v$. Moreover, $x_v>0$ by \eqref{eq:xv_def}, and the singular locus at the equator is located at $x_s=0$ according to \eqref{eq:xs_def}. Thus, we obtain $x_G>x_v>0=x_s$.
\end{proof}
%______________________________________________________________________________________________________
\subsection{Wormhole Throat}
%______________________________________________________________________________________________________
%{\it Wormhole Throat}. 
The throat is defined as the minimum \eqref{eq:areal_min_formal} of the areal function given by \eqref{eq:Areal_formal}, then
\begin{align}\label{Funcion Areal}
\mathcal{A}(x)&=2\pi (L_\pm)^2 \int_{1}^{-1} \frac{e^{k}}{f} \sqrt{\frac{x^2 \pm y^2}{x^2\pm 1}}\, dy.
\end{align}
To determine the location of the throat, we employ numerical methods to obtain the value $x = x_G$.
The numerical analysis provides important insights. First, wormholes do not form for $\lambda_0 \geq 1$. In contrast, when $\lambda_0 \leq 10^{-3}$, $\tau_0 \leq 10^{-4}$, $L_{+} = 2$, and $k_0 = 3/4$ (dilatonic field in superstring theory), they are stable and exhibit a throat in the SU-E case (upper sign). However, in the S-E case (lower sign), the throat is always confined to the region $|x_G|<1<x_v$, then it does not exist. Thus, if we interpret the sub-extreme configuration as a wormhole, its throat is hidden behind the event horizon, and the wormhole geometry only appears in the super-extreme regime. In other words,the SU-E configuration corresponds to a wormhole (WH), while the S-E configuration corresponds to a black hole (BH).
%**********************************************
\subsubsection{S-E case}
%**********************************************
\paragraph{Why no throat in the sub-extreme scenario, but a throat may form in the super-extreme scenario.}
In the S–E regime, any wormhole throat that is intrinsically characterized as a strict minimizer of the areal functional must be located within the regular, chronology-safe domain ($x>x_v$). Because in this branch one has $x_v>1$, the associated variational problem is restricted to an outer region with $x=\mathcal O(1)$, where the conformal factor $e^{k(x,y)}$ stays smooth and bounded and thus cannot produce a geometric barrier capable of generating an interior basin (i.e.\ a genuine critical point) for the area. Consequently, the areal functional admits no acceptable interior minimum, and no regular throat can arise. In contrast, in the SU–E regime the regularity threshold can lie at $x\ll 1$, so the minimization extends into the inner region where $e^{k}$ may increase steeply; this pronounced growth can induce a basin structure for the areal functional and thereby permit a genuine areal minimum to form, yielding an intrinsic throat.
%**********************************************
\subsubsection{SU-E case}
%**********************************************
Numerically, taking $\lambda_0 = 10^{-3}$, $L_{+}=1$, and $k_0=3/4$ (dilatonic field in superstring theory), in the SU-E case wormholes do not appear for $\lambda_0 \geq 1$. In contrast, within the SU-E configuration the throat is located at $x_G = 0$ for $y \in [-1,-45\lambda_0) \cup (45\lambda_0,1]$ (approximately). For $y \in [-45\lambda_0,45\lambda_0]$, the throat moves to a position with $x_G \neq 0$, where $x_G$ now depends on the angle $y_0$ (for instance, for $y_0=0$, $x_G \approx 0.0293>x_v\approx \lambda_0 = 10^{-3}$), see Figure \ref{fig:ExpliGarganta}. Consequently, the SU-E setup is the only one that can be consistently interpreted as a wormhole, with the throat situated at the minimum of the areal function $\mathcal{A}(x)$ satisfying numerically Conjecture \ref{conj:WCCC_topological}.

Although $\mathcal{A}(x)$ is defined through an integral over $y$, the strategy is to minimize the equatorial areal density
\begin{align}\label{eq:aeq_def}
\mathfrak a(x,y)&=\sqrt{g_{yy}(x,y)\,g_{\varphi\varphi}(x,y)} \qquad x>x_v, \nonumber \\
&= \frac{L_+^2}{f_0}\,e^{k(x,y)}\,\sqrt{\mathfrak Q(x,y)},
\end{align}
where $L_{+}^2/f_0 \ >0$, and 
\[
\label{eq:Q_def_y0}
\mathfrak Q(x,y)=(x^2+y^2)(x^2+1)-\lambda_0^2\,\frac{x^2(1-y^2)}{x^2+y^2}.
\]
Because $\mathfrak Q(0,y)=y^2>0$, there is some $\delta(y)>0$ such that $\mathfrak Q(x,y)>0$ whenever $|x|<\delta(y)$. Consequently, $\mathfrak a(\cdot,y)$ is twice continuously differentiable with respect to $x$ in a neighborhood of $x=0$. We now introduce
\begin{equation*}
\mathfrak F(x,y)=\ln\mathfrak a(x,y)=k +\frac12\ln \mathfrak Q,
\end{equation*}
so that $\mathfrak a_x=\mathfrak a\,F_x$, $\mathfrak a>0$ near $x=0$, and $\partial_x \mathfrak \, (\cdot) =(\cdot)_{,x}$.
Differentiating
{\footnotesize
\begin{align}\label{eq:Fx_simplified}
\mathfrak F_x
&=
\frac{k_{0}\lambda_0^{2}\,x\,(1-y^{2})
\Big(9x^{4}y^{2}-x^{4}-6x^{2}y^{4}+10x^{2}y^{2}+y^{6}-5y^{4}\Big)}
{(x^{2}+y^{2})^{5}}
\nonumber\\
&\quad+
\frac{x\Big(\lambda_0^{2}y^{2}(y^{2}-1)+(x^{2}+y^{2})^{2}(2x^{2}+y^{2}+1)\Big)}
{(x^{2}+y^{2})\Big((x^{2}+1)(x^{2}+y^{2})^{2}-\lambda_0^{2}x^{2}(1-y^{2})\Big)}.
\end{align}
}
The simplest root of $\mathfrak F_x = 0$ is $x = 0$, and thus from $\mathfrak F_x(0,y) = 0$ it follows that $\mathfrak a_x (0,y) = 0$. In order to confirm that this point corresponds to a WH throat, we must require $\mathfrak a_{xx}>0$, taking into account that $\partial_x a_{x}=\mathfrak a(\mathfrak F_{xx}+ \mathfrak F_x^2)$ and $\mathfrak a$ is positive in a neighborhood of $x=0$, the sign of $\mathfrak a(0,y)$ is determined by the sign of $(\mathfrak F_{xx}(0,y)+ \mathfrak F_x^2(0,y))$, which simplifies to $\mathfrak F_{xx}(0,y)$.

The complete expression evaluated at the point $x=0$ and $y\neq 0$ is:
{\footnotesize
\begin{equation*}
    \mathfrak F_{xx}(0,y)=
\frac{1+y^2}{y^2}
-\lambda_0^2\,\frac{1-y^2}{y^4}
-k_0\lambda_0^2\,\frac{(1-y^2)(5-y^2)}{y^6}.
\end{equation*}
}
There are two important things in the region $y_0\in(0,1)$, 
\[
\lim_{y\to 1^-}\mathfrak F_{xx}(0,y)=\lim_{y\to 1^-}\frac{1+y^2}{y^2}=2>0,
\]
and as $y\to 0^+$, the last term in $\mathfrak F_{xx}(0,y)$ dominates:
\[
-k_0\lambda_0^2\frac{(1-y^2)(5-y^2)}{y^6}\sim -5k_0\lambda_0^2\,y^{-6}\to -\infty,
\]
hence
\[
\lim_{y\to 0^+}\mathfrak F_{xx}(0,y)=-\infty.
\]
Since $\mathfrak F_{xx}(0,y)$ is continuous on $(0,1)$ and changes sign, it indicates that there exists at least one $y_1\in(0,1)$ such that $\mathfrak F_{xx}(0,y_1)=0$, or equivalently 
\[
(1+y_1^2)\,y_1^4 = \lambda_0^2(1-y_1^2)\,y_1^2 + k_0\lambda_0^2(1-y_1^2)(5-y_1^2).
\]
When the region of displacement is restricted near the equatorial plane, such that $y_0 \ll 1$, then
\[
(1+y_1^2)y_1^4\simeq y_1^4,
\qquad
(1-y_1^2)\simeq 1,
\qquad
(5-y_1^2)\simeq 5.
\]
Then $\mathfrak F_{xx}(0,y)$ reduces to the quadratic equation in $z=y_1^2$,
\[
z^2\simeq \lambda_0^2 z+5k_0\lambda_0^2.
\]
Solving,
\[
z\simeq \frac{\lambda_0^2+\sqrt{\lambda_0^4+20k_0\lambda_0^2}}{2}.
\]
For $\lambda_0\ll 1$, the square root is dominated by $\sqrt{20k_0}\,\lambda_0$, consequently, the throat location structure is:
\begin{subequations}
\begin{align}\label{eq:GARGANTAx0_belt}
&x_G(y)=0 \qquad &&\text{for}\quad |y|\ge y_1,\\
&x_G(y)\neq 0 \qquad &&\text{for}\quad |y|< y_1,
\end{align}
\end{subequations}
with $\lambda_0\ll 1, \ k_0>0$ and
\begin{equation}\label{eq:y1 limiteGargantax0}
y_1\sim (5k_0)^{1/4}\,\lambda_0^{1/2}.
\end{equation}
If we use $\lambda_0=10^{-3}$ and $k_0=3/4$, we obtain $y_1 \approx 0.0440056$.
%....................................................................................
\paragraph{Equatorial plane.}
%....................................................................................
The metric components evaluated on the equatorial plane are
\[
g_{yy}(x,0)=\frac{L_+^2}{f_0}\,x^2\,e^{2k(x,0)},
\qquad
k(x,0)=\frac{k_0\lambda_0^2}{4x^4},
\]
and
\[
g_{\varphi\varphi}(x,0)=\frac{L_+^2}{f_0}\,\frac{x^4+x^2-\lambda_0^2}{x^2}.
\]
Therefore,
\[
\mathfrak a_{\mathrm{eq}}(x)
=\frac{L_+^2}{f_0}\,e^{k(x,0)}\,\sqrt{x^4+x^2-\lambda_0^2},
\qquad x>x_v.
\]
Since the prefactor $L_+^2/f_0$ is positive constant, the critical points are determined by minimizing
\[
\widetilde{\mathfrak a}_{\mathrm{eq}}(x)=e^{k(x,0)}\,\sqrt{x^4+x^2-\lambda_0^2}.
\]
Once more, it is more practical to work with $\ln{\mathfrak a}_{\mathrm{eq}}$, namely,
\[
\ln \widetilde{\mathfrak a}_{\mathrm{eq}}(x)=k(x,0)+\frac12\ln(x^4+x^2-\lambda_0^2),
\]
then using $k'(x,0)=-k_0\lambda_0^2/x^5$ on $\frac{d}{dx}\ln \widetilde{\mathfrak a}_{\mathrm{eq}}(x)=0$ we obtain
\[
\frac{2x^3+x}{x^4+x^2-\lambda_0^2}=\frac{k_0\lambda_0^2}{x^5},
\qquad x>x_v.
\]
Expanding the expression yields the explicit polynomial equation
{\small
\begin{equation}\label{eq:critical_poly_x}
2x^8+x^6-k_0\lambda_0^2x^4-k_0\lambda_0^2x^2+k_0\lambda_0^4=0,
\qquad x>x_v>0.
\end{equation}
}
\paragraph{Approximation $x_G \ll 1$, $\lambda_0 \ll 0$ and $k_0>0$.}
As a first approximation, we will employ the full expression of the critical-point equation
\[
    \frac{k_0\lambda_0^2}{x^5}=\frac{2x^3+x}{x^4+x^2-\lambda_0^2} \approx \frac{1}{x},
\]
where $\lambda_0 \ll x$, then the WH at the equatorial plane is
\begin{equation}\label{eq:GARGANTA plano ecuatorial}
    x_G^{(\mathrm{eq})} \approx (k_0)^{1/4}\,\lambda_0^{1/2}.
\end{equation}
If we use $\lambda_0=10^{-3}$ and $k_0=3/4$, we obtain $x_G^{(\mathrm{eq})} \approx 0.0294283$.
%....................................................................................
\paragraph{Positivity of the second derivative at equatorial plane.}
%....................................................................................
Define $\mathfrak  D(x)=x^4+x^2-\lambda_0^2$, and $\mathfrak F(x)=\ln \widetilde{\mathfrak a}_{\mathrm{eq}}(x)$, therefore
\begin{equation}\label{eq:Fsecond_exact}
\mathfrak F''(x)=k''(x,0)+\frac12\left(\frac{\mathfrak D''(x)}{\mathfrak D(x)}-\frac{(\mathfrak D'(x))^2}{\mathfrak D(x)^2}\right),
\end{equation}
where
{\footnotesize
\begin{equation*}
k''(x,0)=\frac{5k_0\lambda_0^2}{x^6},
\qquad
\mathfrak D'(x)=4x^3+2x,
\qquad
\mathfrak D''(x)=12x^2+2.
\end{equation*}
}
Now apply the same small‑$x$ equatorial domain:
{\footnotesize
\[
x\ll 1,\qquad x\gg\lambda_0 \ \text{implies}\ D(x)\sim x^2,\quad D'(x)\sim 2x,\quad D''(x)\sim 2.
\]
}
Plugging this into \eqref{eq:Fsecond_exact} yields the leading behavior
\[
F''(x)\sim \frac{5k_0\lambda_0^2}{x^6}
+\frac12\left(\frac{2}{x^2}-\frac{(2x)^2}{x^4}\right)
=
\frac{5k_0\lambda_0^2}{x^6}-\frac{1}{x^2}.
\]
Finally, using $x^4\sim k_0\lambda_0^2$. Writing $x^6=x^2x^4\sim x^2(k_0\lambda_0^2)$ yields
\[
\frac{5k_0\lambda_0^2}{x^6}\sim \frac{5k_0\lambda_0^2}{x^2(k_0\lambda_0^2)}=\frac{5}{x^2}.
\]
Hence,
\[
F''\!\left(x_G^{(\mathrm{eq})}\right)\sim \frac{5}{x^2}-\frac{1}{x^2}
=\frac{4}{x^2}>0.
\]
Therefore, this configuration corresponds to the widest WH throat and complies with the WCCC, i.e. $x_G^{(\mathrm{eq})}>x_v>x_s=0$.
%
%______________________________________________________________________________________________________
\subsection{WCCC examined for our exact SU-E family of dilatonic wormholes}
\label{subsec:apply_SUe_formalism}
%______________________________________________________________________________________________________
\paragraph{Defect set}
In our situation, the ring defect is located at $(x,y)=(0,0)$,  then $x_s(0)=0$. Examining CTCs through $\chi = g_{\varphi\varphi}$ identifies a chronology boundary, denoted by $x_v(y)$, at the equatorial plane $x_v(0)>x_s(0)=0$.
In the wormhole regime we are considering, no event horizon is present, so $\mathcal V_{\rm H}=\emptyset$, and $x_h(y)$ can be omitted. Consequently, the unified censorship radius at $y=0$ is
\[
x_\ast(0)=\max\{x_s(0),x_v(0)\}=x_v(0).
\]
\paragraph{Throat location and enclosure inequality.}
Independently of this consideration, the throat is intrinsically defined as the unique strict minimizer of the areal functional $\mathcal A(x)$ in \eqref{eq:Areal_formal}, which determines a throat profile $x_G(y)$.
Our explicit estimates and inequalities show that $x_\ast(y) < x_G(y)$ over the entire relevant region and, in particular, at $y = 0$. Consequently, Proposition~\ref{prop:operational_enclosure} is applicable, implying that the throat completely encloses the entire defect set in its interior. Hence, our solution on the equatorial plane satisfies Conjecture~\ref{conj:WCCC_topological}. Moreover, numerical evidence shows that all space-time pathologies are concealed by the deformed throat; see Figure~\ref{fig:ExpliGarganta}.
%
%
%______________________________________________________________________________________________________
%\subsection{The Cartan-Penrose diagram}
%______________________________________________________________________________________________________
%{\it The Cartan-Penrose diagram}. 
Following \cite{Chrusciel:2020fql}, we now restrict our attention to the wormhole (SU-E) dilatonic case with $k_0>0$, and we take $\varphi=\varphi_0$ to be constant. Under these assumptions, \eqref{ds sp} ($ct \rightarrow t$) reduces to
{\small
\begin{align}\label{ds sin varphi}
    &ds^2=-f dt^2 +f^{-1} e^{2k} (d\rho ^2 + dz^2), \notag \\
    &=-f dt^2 +\frac{e^{2k}}{f}(L_{+})^2(x^2+y^2) \left( \frac{dx^2}{x^2+1} + \frac{dy^2}{1-y^2}\right),
\end{align}
}
and we focus on the hypersurface $y=y_0$ with $y_0$ constant to analyze the corresponding Penrose diagrams, then $ds^2=-f\,dt^2+\mathfrak G\,dx^2$, where $f=f_0$, and
\begin{equation}\label{eq:G_Garigoleada}
\mathfrak G(x;y_0) =\frac{L_+^2}{f_0}\,e^{2k(x,y_0)}\,\frac{(x^2+y_0^2)}{x^2+1}.
\end{equation}
%......................................
\paragraph{Manifold regularity vs chronology regularity.}
%......................................
Throughout, we maintain a clear distinction between \emph{manifold regularity} and \emph{chronological regularity}. By \emph{manifold regularity} we refer solely to the requirement that, in the chosen chart, the metric components form a smooth Lorentzian tensor field. For the present class of space-times, this condition is violated only on the ring-defect locus, whose projection onto the $(x,y)$-plane is the single point $(x,y)=(0,0)$. Consequently, we take the physical manifold to be represented by the coordinate patch
\begin{equation}\label{eq:manifold_regular_domain}
\mathcal M_{\rm reg}^{\rm (man)}:\quad (x,y)\in\mathbb{R}^2\setminus\{(0,0)\}.
\end{equation}
In contrast, \emph{chronological regularity} requires, in addition, the absence of CTCs, which, in our stationary, axisymmetric framework, is governed by the sign of $g_{\varphi\varphi}$. Consequently, for each fixed value $y=y_0$, there exists a threshold $x_*(y_0)\ge 0$ such that the chronology-safe region is
\begin{equation}\label{eq:chronology_regular_domain}
\mathcal M_{\rm reg}^{\rm (chr)}(y_0):\quad x>x_*(y_0).
\end{equation}
Thus, the condition $x > x_*(y_0)$ is strictly stronger than mere manifold regularity: it rules out the CTC belt, even though the manifold itself is already well-defined everywhere except at $(0,0)$.
%......................................
\paragraph{Regularity domain on a fixed $y_0$-slice (manifold regularity).}
%......................................
Within the SU-E family, the sole non-manifold point in the $(x,y)$-plane is the ring defect at $(0,0)$. Therefore, for a fixed slice $y = y_0$, the manifold-regular domain in $x$ is
\[
I_{y_0}=
\begin{cases}
\mathbb{R}, & y_0\neq 0,\\
\mathbb{R}\setminus\{0\}=(-\infty,0)\cup(0,\infty), & y_0=0.
\end{cases}
\]
On $I_{y_0}$ the function $k(x,y_0)$ is real and finite, so $e^{2k(x,y_0)}>0$. Since also $x^2+y_0^2>0$ and $x^2+1>0$, we obtain
\[
\mathfrak G(x;y_0)>0\qquad \text{for all}\qquad  x\in I_{y_0}.
\]
%**********************************************
\subsubsection{Tortoise coordinate}
%**********************************************
For every $y_0$-slice, the tortoise variable is defined as
{\small
\begin{align}\label{Variable Tortuga}
    f(d\,l_{y0}){}^2 & \equiv \mathfrak G\,dx^2, \text{implies} \frac{d\,l_{y0}}{dx}=\sqrt{\frac{\mathfrak G}{f}}=\frac{L_+}{f_0}\,e^{k(x,y_0)}\sqrt{\frac{x^2+y_0^2}{x^2+1}}. 
    %\notag \\ & = \frac{e^{2k}}{f} L^2 (x^2+y^2) \left( \frac{dx^2}{x^2+1} + \frac{dy^2}{1-y^2}\right),
\end{align}
}
since $\mathfrak G$ is strictly positive and regular, then  
\[
\frac{d\,l_{y0}}{dx}>0, \quad \text{for all}\qquad  x\in I_{y_0} ,
\]
Since $\frac{d\,l_{y0}}{dx}$ is continuous on the compact interval $[x_v,x_G]$ at $y_0$, the Extreme Value Theorem implies that $\frac{d\,l_{y0}}{dx}$ attains its minimum there. Moreover, since $\frac{d\,l_{y0}}{dx}>0$ on $[x_v,x_G]$, this minimum is strictly positive:
\[
\mathfrak m=\min_{x\in[x_v,x_G]}\frac{d\,l_{y0}}{dx}>0.
\]
Hence
\begin{equation*}
\int_{x_v}^{x_G}\frac{d\,l_{y0}}{dx}\,dx
\;\ge\;
\int_{x_v}^{x_G} \mathfrak m\,dx
\;=\;
\mathfrak m \,(x_G-x_v)
\;>\;0,
\end{equation*}
then by the Fundamental Theorem of Calculus, $\int_{x_v}^{x_G}\frac{dl}{dx}\,dx=l(x_G)-l(x_v)$
\begin{equation}\label{eq:lG_Mayora_lv}
    l(x_G)>l(x_v).
\end{equation}
%
%......................................
\paragraph{Tortoise coordinate of the ring singularity.}
%......................................
On the equatorial plane $y=0$ the dilatonic SU-E function $k$ satisfies
\begin{equation*}
k(x,0)=\frac{k_0\lambda_0^2}{4x^4},
\qquad k_0>0,
\end{equation*}
so in the fixed-$y_0$ sector the tortoise coordinate is defined by
{\small
\begin{equation}\label{eq:dl_dx_equator_minusinf}
\frac{dl}{dx}
=
\sqrt{\frac{\mathfrak G(x;0)}{f_0}}
=
\frac{L_+}{f_0}\exp\!\left(\frac{k_0\lambda_0^2}{4x^4}\right)\frac{x}{\sqrt{x^2+1}},
\qquad x>0.
\end{equation}
}
In particular,
\begin{equation*}
\frac{dl}{dx}>0 \qquad \text{for all}\qquad  x>0,
\end{equation*}
which implies that $l$ is strictly increasing over the entire right-hand interval $(0,+\infty)$.
Fix any $x_v>0$. Then, for $x$ in the interval $(0,x_v]$ we have
\[
x^2+1\le x_v^2+1
\qquad\text{implies}\qquad
\frac{1}{\sqrt{x^2+1}}\ge \frac{1}{\sqrt{x_v^2+1}}.
\]
Multiplying by the nonnegative factor $x\ge0$ yields
\begin{equation}
\label{eq:x_over_sqrt_bound}
\frac{x}{\sqrt{x^2+1}}
\;\ge\;
\frac{x}{\sqrt{x_v^2+1}}
\qquad \text{for all }x\in(0,x_v].
\end{equation}
Substituting \eqref{eq:x_over_sqrt_bound} into \eqref{eq:dl_dx_equator_minusinf} gives the inequality
{\small
\begin{equation}\label{eq:dl_dx_lower_bound}
\frac{dl}{dx}
\ge
\frac{L_+}{f_0\sqrt{x_v^2+1}}\,
x\,
\exp\!\left(\frac{k_0\lambda_0^2}{4x^4}\right)
\qquad \text{for all }x\in(0,x_v].
\end{equation}
}
For any $0<\varepsilon<x_v$, integrate \eqref{eq:dl_dx_lower_bound} from $\varepsilon$ to $x_v$:
\begin{align*}
\int_{\varepsilon}^{x_v}\frac{dl}{dx}\,dx
&\ge
\frac{L_+}{f_0\sqrt{x_v^2+1}}
\int_{\varepsilon}^{x_v} x\,
\exp\!\left(\frac{k_0\lambda_0^2}{4x^4}\right)\,dx.
\end{align*}
By the Fundamental Theorem of Calculus, $\int_{\varepsilon}^{x_v}\frac{dl}{dx}\,dx=l(x_v)-l(\varepsilon)$, yields
\begin{equation*}
l(x_v)-l(\varepsilon)
\ge
\frac{L_+}{f_0\sqrt{x_v^2+1}}
\int_{\varepsilon}^{x_v} x\,
\exp\!\left(\frac{k_0\lambda_0^2}{4x^4}\right)\,dx.
\end{equation*}
Now, using the change of variable $u=\frac{a}{x^4}$ where $a=\frac{k_0\lambda_0^2}{4}>0$, we transform 
\[
\int_{\varepsilon}^{x_v} x\,e^{a/x^4}\,dx=\frac{\sqrt a}{4}\int_{a/x_v^4}^{a/\varepsilon^4} u^{-3/2}\,e^{u}\,du.
\]
As $\varepsilon\to0^+$ we have $a/\varepsilon^4\to+\infty$, and thus the aforementioned integral becomes
{\small
\[
\frac{\sqrt a}{4}\int_{a/x_v^4}^{a/\varepsilon^4} u^{-3/2}\,e^{u}\,du
\xrightarrow[\varepsilon\to0^+]{}\;
\frac{\sqrt a}{4}\int_{a/x_v^4}^{+\infty} u^{-3/2}\,e^{u}\,du.
\]
}
Next, recall that $x_v \ll 1$, which implies $U_v=\frac{a}{x_v^4} \gg 1$. Consequently, we introduce the function
\[
\Phi(u)=\frac{u}{2}-\frac{3}{2}\ln u,\qquad u>0.
\]
Thus,
\[
\Phi'(u)=\frac12-\frac{3}{2u}=\frac{u-3}{2u},
\]
thus, $\Phi$ is strictly increasing for all $u>3$. In addition, since $\Phi(u)\to +\infty$ as $u\to +\infty$ (because $\ln u / u \to 0$ when $u\to\infty$, so that
\begin{equation*}
    \Phi(u)=\frac{u}{2}\left(1-3\frac{\ln u}{u}\right)\xrightarrow[u\to\infty]{}+\infty,
\end{equation*}
it follows that there exists some $U_\star>3$ such that $\Phi(u)\ge 0$ for every $u\ge U_\star$. The inequality $\Phi(u)\ge 0$ is exactly
{\small
\[
\frac{u}{2}\ge \frac{3}{2}\ln u
\quad\text{if and only if}\quad
e^{u/2}\ge u^{3/2}\]
\[
\quad\text{if and only if}\quad
u^{-3/2}e^{u}\ge e^{u/2},
\]
}
Thus, in our situation we have $U_v \gg 3$, and therefore we may apply the inequality $u^{-3/2} e^{u} \ge e^{u/2}$. To conclude, by applying the proposed inequality, we obtain
{\small
\[
\int_{U_v}^{a/\varepsilon^4}u^{-3/2}e^{u}\,du
\ \ge\
\int_{U_v}^{a/\varepsilon^4}e^{u/2}\,du
=
2\Big(e^{\frac{a}{2\varepsilon^4}}-e^{\frac{U_v}{2}}\Big),
\]
\[
\text{therefore}\quad \int_{\varepsilon}^{x_v}x\,e^{a/x^4}dx
\xrightarrow[\varepsilon\to0^+]{}+\infty
\]
}
Because $l(x_v)$ is a fixed finite number, the only way the difference can diverge to $+\infty$ is that
\begin{equation}\label{eq:lv_Mayora_ls}
l(\varepsilon)\xrightarrow[\varepsilon\to0^+]{}-\infty<l(x_v).
\end{equation}
%**********************************************
\subsubsection{Penrose compactification}
%**********************************************
We now perform the standard 2D conformal compactification of the $(t,x)$ sector on a fixed slice $y=y_0$ (and $\varphi=\varphi_0$), assuming the SU-E regime with $f=f_0>0$ constant. The metric takes the form 
\begin{equation}
\label{eq:2D_sector_general}
ds^2=-f_0\,dt^2+f_0\,dl_{y_0}^2,
\end{equation}
where the tortoise coordinate $l_{y_0}=l_{y_0}(x)$.
%......................................
\paragraph{Null coordinates and flat conformal form.}
%......................................
Introduce the radial null coordinates
\begin{equation}\label{eq:uv_def_penrose}
u=t-l_{y_0},\qquad v=t+l_{y_0},
\end{equation}
so that
\begin{equation}\label{eq:ds_uv_form}
ds^2_{(u,v)}=-f_0\,du\,dv.
\end{equation}
This is conformally equivalent to $2$D Minkowski space, characterized by a constant conformal factor $f_0$.
%......................................
\paragraph{Compactification.}
%......................................
Let
{\small
\begin{equation}\label{eq:UV_def_penrose}
u=\tan U,\qquad v=\tan V,
\qquad U,V\in\left(-\frac{\pi}{2},\frac{\pi}{2}\right),
\end{equation}
}
so that $du=\sec^2U\,dU$ and $dv=\sec^2V\,dV$, and therefore
\begin{equation}\label{eq:ds_UV}
ds^2_{(U,V)}=-f_0\,\sec^2U\,\sec^2V\,dU\,dV.
\end{equation}
With the conformal factor
\begin{equation}\label{eq:Omega_def}
\mathbf \Omega=\cos U\,\cos V,
\end{equation}
the rescaled metric $\bar s^2=\mathbf \Omega^{2}\,ds^2$ becomes
\begin{equation}\label{eq:dsbar_UV}
d\bar s^2=-f_0\,dU\,dV.
\end{equation}
Finally let
\begin{subequations}
\begin{align}
&\mathbf T=\frac{V+U}{2},\qquad \mathbf R=\frac{V-U}{2} \\
&\text{implies} \qquad
V=\mathbf T+\mathbf R,\ \ U=\mathbf T-\mathbf R,\label{eq:TR_def}
\end{align}
\end{subequations}
the metric transforms to
\begin{equation}\label{eq:dsbar_TR}
d\bar s^2=-f_0\,(d\mathbf T^2-d\mathbf R^2).
\end{equation}
%......................................
\paragraph{Conformal boundary.}
%......................................
The conformal boundary is given by $\mathbf \Omega=0$, that is, by the condition $\cos U=0$ or $\cos V=0$, or equivalently
\[
U=\pm\frac{\pi}{2}\quad\text{or}\quad V=\pm\frac{\pi}{2}.
\]
Using \eqref{eq:TR_def}, these are the four null lines
{\small
\begin{align}\label{eq:Omega0_lines}
U=\pm\frac{\pi}{2}\ &\Longleftrightarrow\ \mathbf T-\mathbf R=\pm\frac{\pi}{2}
\ \Longleftrightarrow\ \mathbf T=\pm\frac{\pi}{2}+\mathbf R,\\
V=\pm\frac{\pi}{2}\ &\Longleftrightarrow\ \mathbf T+\mathbf R=\pm\frac{\pi}{2}
\ \Longleftrightarrow\ \mathbf T=\pm\frac{\pi}{2}-\mathbf R,
\end{align}
}
represents the \emph{future/past null infinity} $\mathscr I^\pm$ and the time-like infinities $i^{\pm}$ in the usual Penrose diagram, but only for the universe $\mathcal{U}_1$. We must repeat the same construction for the universe $\mathcal{U}_2$, obtaining an analogous result but with the replacement $x \rightarrow -x$.

%--------------------------------------------------------
\paragraph{Ordered positions.}
In this paragraph, we examine the variable $\mathbf R$ on the equatorial plane, specified by $y=0$. Suppose $x_G>x_v>x_s$, then, as shown in the previous subsection, we have $l(\varepsilon)\xrightarrow[\varepsilon\to x_s^+]{}-\infty<l(x_v)<l(x_G)$. We now define the corresponding compactified radii at $t=0$ by
\begin{equation}\label{eq:Rv_RG_defs}
\mathbf R_v=\arctan(l_v),\qquad \mathbf R_G=\arctan(l_G).
\end{equation}
Since $\arctan$ is strictly increasing on $\mathbb R$, $l(\varepsilon)\xrightarrow[\varepsilon\to x_s^+]{}-\infty<l(x_v)<l(x_G)$ yields
\begin{equation}\label{eq:R_order_full}
R_s=-\frac{\pi}{2}\;<\;R_v\;<\;R_G\;<\;\frac{\pi}{2}.
\end{equation}
This is exactly the sequence of inequalities required to construct the Carter–Penrose diagram featuring three distinguished vertical (constant‑$R$) lines: the defect boundary or ring singularity at $R_s$, the chronology bound at $R_v$, and the throat at $R_G$, with the throat located strictly outside the onset of chronology violation in the compactified picture.

%--------------------------------------------------------
\paragraph{Causal inaccessibility of the ring singularity.}
%--------------------------------------------------------
Using the variables defined in \eqref{eq:uv_def_penrose}, the metric can be written as $ds^2_{(u,v)} = -f_0\,du\,dv$. More importantly, radial null geodesics are described by $u = \mathrm{const}$ for ingoing rays and $v = \mathrm{const}$ for outgoing rays, and we know that $l(\varepsilon)\xrightarrow[\varepsilon\to x_s^+]{}-\infty$. Consequently, along an ingoing null ray $u = u_0$ one finds $t = u_0 + l(\varepsilon)\to -\infty$ as $\varepsilon\to x_s^+$, whereas along an outgoing null ray $v = v_0$ one has $t = v_0 - l(\varepsilon)\to +\infty$ as $\varepsilon\to x_s^+$. Hence, the point $\varepsilon = x_s$ can only be approached at infinite coordinate time and cannot be reached from any event with finite $t$ by a future-directed causal curve.

Thus, the ring singularity does not belong to the physical spacetime manifold, it corresponds instead to an ideal endpoint residing on the conformal boundary (a causal limit point), rather than to an actual spacetime event that can be attained causally.

%%%%%%%%%%%%%%%%%%%%%%%%%%%%%%%%%%%%%%%%%%%%%%%%%%%%%%%%%%%%%%%%%%%%%%%%%%%%%%%%%%%%%%%%%%%%%%%%%%%
%%%%%%%%%%%%%%%%%%%%%%%%%%%%%%%%%%%%%%%%%%%%%%%%%%%%%%%%%%%%%%%%%%%%%%%%%%%%%%%%%%%%%%%%%%%%%%%%%%%
\section{Plotting}
%%%%%%%%%%%%%%%%%%%%%%%%%%%%%%%%%%%%%%%%%%%%%%%%%%%%%%%%%%%%%%%%%%%%%%%%%%%%%%%%%%%%%%%%%%%%%%%%%%%
%%%%%%%%%%%%%%%%%%%%%%%%%%%%%%%%%%%%%%%%%%%%%%%%%%%%%%%%%%%%%%%%%%%%%%%%%%%%%%%%%%%%%%%%%%%%%%%%%%%
%============================================================
All curves in the Penrose diagram are represented in conformal coordinates $(\mathbf T,\mathbf R)\in(-\pi/2,\pi/2)\times(-\pi/2,\pi/2)$, employing a common arctangent mapping. Given a real parameter $t\in\mathbb{R}$ and a fixed width parameter $l>0$, we introduce
\begin{align*}
\mathbf R(t;l)&=\frac12\!\left[\arctan(t-l)-\arctan(t+l)\right],\\
\mathbf T(t;l)&=\frac12\!\left[\arctan(t-l)+\arctan(t+l)\right],
\end{align*}

To illustrate the throat and the bound at which causal violation begins, we choose the parameters $\mathbf R( l_G=0.55) \ \ > \ \ \mathbf R( l_v=0.25)$ at the axis $\mathbf T=0$, emphasizing that this choice serves purely demonstrative purposes and consider $t \in (-80,80)$ for the universe 1. In the corresponding alternate universe, we simply apply $l \rightarrow -l$.

For the traveler who stays within their own universe, the same procedure was applied, but with the parameter $\mathbf R(l_0=-\pi /4)$ on the axis $\mathbf{T} =0$ for universe 1; for universe 2, the transformation $l \rightarrow -l$ was once again employed.

The crossing traveler is obtained by performing a Lorentz-like rescaling of the arctangent’s arguments, controlled by a parameter $v_{\rm cross}\in(0,1)$:
{\footnotesize
\begin{align*}
&\mathbf R_{\rm cross}(t;v_{\rm c})=\frac12\!\left[\arctan\!\big((1-v_{\rm c})t\big)-\arctan\!\big((1+v_{\rm c})t\big)\right], \\
&\mathbf T_{\rm cross}(t;v_{\rm c})=\frac12\!\left[\arctan\!\big((1-v_{\rm c})t\big)+\arctan\!\big((1+v_{\rm c})t\big)\right].
\end{align*}
}
The chosen parameters were $v_{\rm cross} = 0.90$, so the final time at which the curves stopped being drawn was $t_{\rm start} = \frac{l_G}{v_{\rm cross}} = \frac{0.55}{0.90}$ for universe 1. For universe 2, we simply applied $l \rightarrow -l$ while keeping the same value of $t_{\rm start}$, which in this case corresponded to the initial time at which the geodesic began to be drawn.

\begin{figure}[h!]
  \centering

  \includegraphics[width=\linewidth,height=0.48\textheight,keepaspectratio]{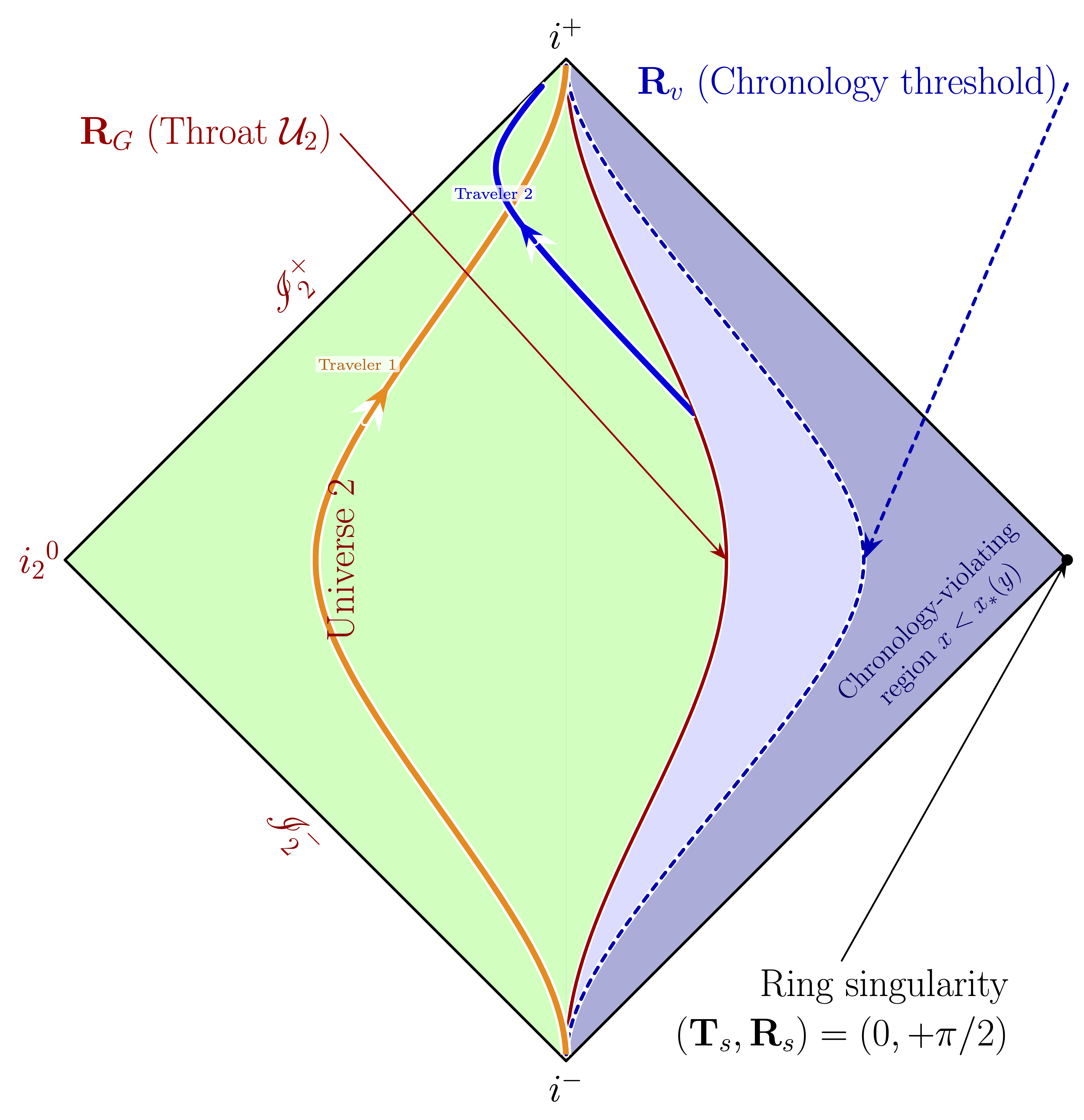}

  \vspace{2mm}

  \includegraphics[width=\linewidth,height=0.48\textheight,keepaspectratio]{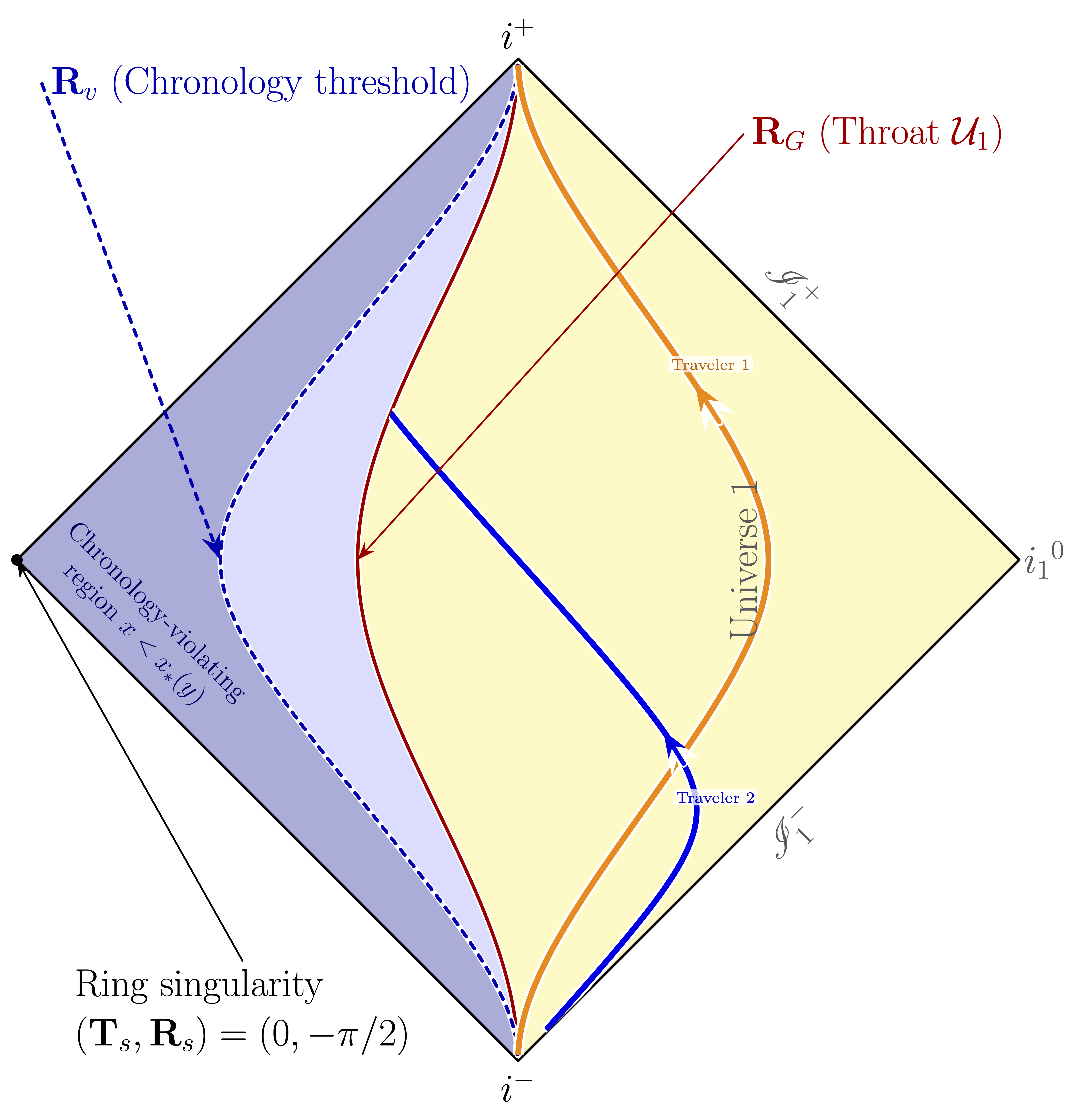}

  \caption{Carter–Penrose diagrams corresponding to each universe at the equatorial plane $y_0=0$.}
  \label{fig:DiagPen Separados}
\end{figure}

%....................%============================================================%============================================================%============================================================
In Figure \ref{fig:DiagPen Separados} we can see the separated Carter–Penrose diagrams for each universe. The upper panel shows the Penrose diagram for universe 2 (green) with two worldlines: an orange trajectory for a traveler who remains in universe 2, and another for a traveler arriving from universe 1 and then staying in universe 2. The dashed purple line marks the boundary of causal violation, the right-hand diamond vertex represents the ring singularity outside the spacetime manifold, and the solid red line indicates the wormhole throat. The lower panel shows the Penrose diagram for universe 1 (yellow), again with two worldlines: an orange trajectory for a traveler who never leaves universe 1, and another for a traveler who starts in universe 1 and heads to universe 2. Here, the dashed purple line marks where causal violation begins, the left-hand diamond vertex represents the ring singularity outside spacetime, and the solid red line indicates the wormhole throat.

\begin{figure}[h!]
  \centering
  \includegraphics[
  width=\linewidth,
  height=\textheight,
  keepaspectratio
]{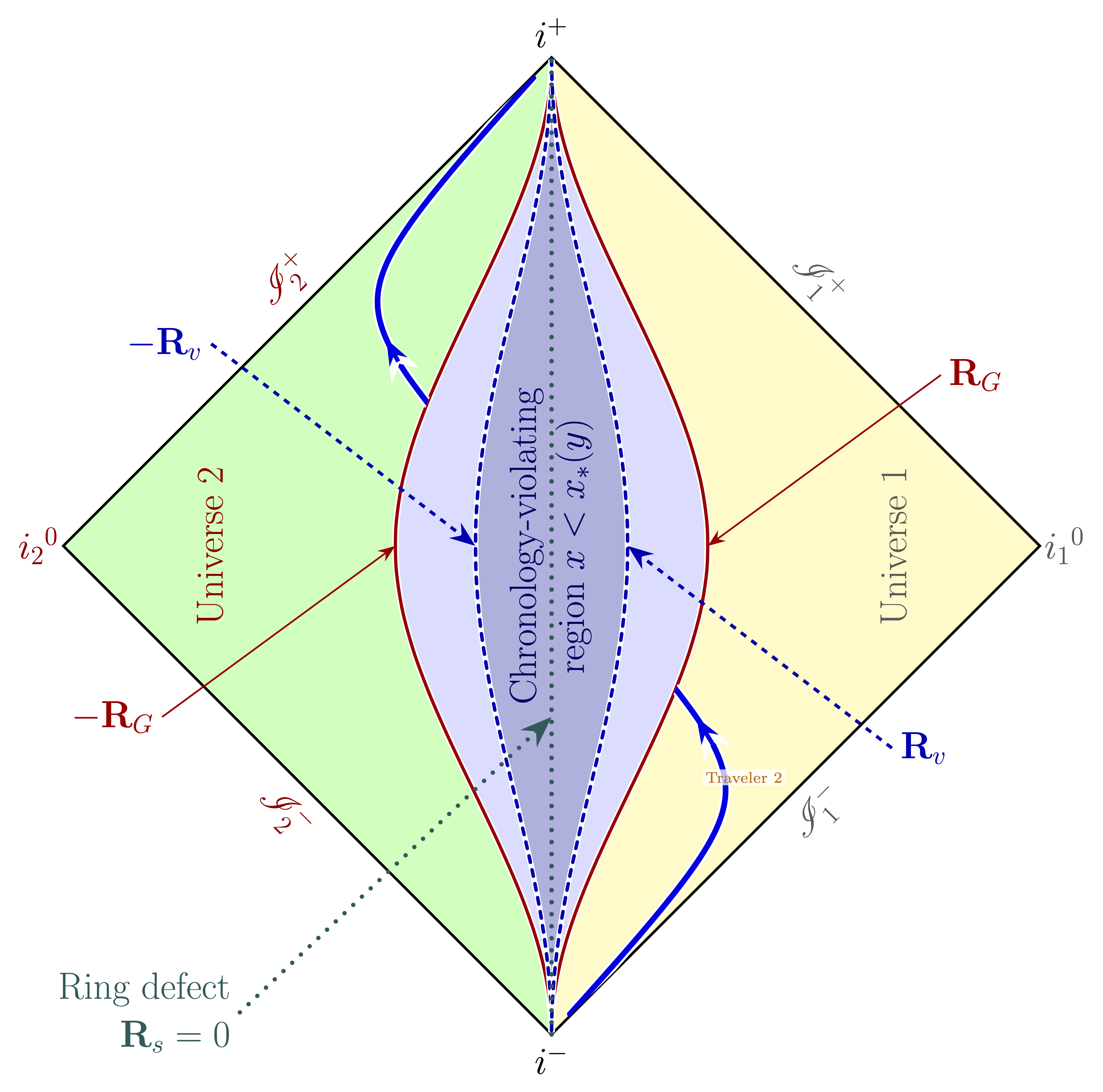}
\caption{Carter-Penrose diagram at the equatorial plane $y_0=0$, representing the WH ring. The dashed lines represent throats topologically connecting $R_G$ to $-R_G$, which cover the singularity of the ring and the chronology-violation region $x<x_*(0)$.}
    \label{fig:EstructuraCausalDiagrama}
\end{figure}

The figure \ref{fig:EstructuraCausalDiagrama} was obtained precisely by identifying the throat \(R_G\) with the throat \(-R_G\) of both universes, that is, the pseudo-spheres \(\mathbb{S}^2\) of radius \(r = l_1\) are topologically identified in Boyer–Lindquist coordinates. In cylindrical coordinates (Weyl-coordinates), the throat is at $z=0$, and the $z=0^{+}$ plane is topologically identified as $z=0^{-}$. If a traveler aims to reach the causality-violating region or the ring singularity, they must first arrive at the throat, and doing so inevitably transfers them into the other universe. The diagram consists of three regions: the green region on the left, representing universe 2, and the yellow region, representing universe 1. Both regions contain a future timelike infinity ($i_+$) and a past timelike infinity ($i_-$), yet these are distinct in universe 1 and universe 2, and spatial infinities, which are denoted by $i_1{}^0$ and $i_2{}^0$. Each universe also has its own past and future null infinities ($\mathscr{I}^{\pm}_{1,2}$). 

The third region, shown in blue, is split into a chronology-violating zone where $g_{\varphi \varphi}<0$ and a regular chronology-preserving zone where $g_{\varphi \varphi}>0$. It is important to note that the ring singularity itself is not part of space-time. In the diagram, it appears as a ring-shaped defect. Likewise, we have drawn a single traversing traveler to illustrate explicitly and schematically how a traveler can start in one universe and emerge in the other; this trajectory is represented by the blue curve.

%------------------------------------------------------------

Thus, no future-directed causal $J$ curve that originates in the exterior region of either end and arrives at $\mathscr I^+_{1,2}$ can pass through $\mathcal V$. In other words, the defect set is causally invisible from both components of null infinity:
\begin{equation}
\label{eq:WCCC_statement}
J^{-}(\mathscr I^+_1)\cap \mathcal V=\emptyset,
\qquad
J^{-}(\mathscr I^+_2)\cap \mathcal V=\emptyset.
\end{equation}
In words: \emph{all curvature or chronology defects are topologically hidden by the throat: an observer coming from either asymptotic region encounters $\mathcal T$ first and is carried over to the opposite exterior, while the defect set inside remains out of reach.}

%------------------------------------------------------------
%\vspace{-6pt}
\section{Conclusions}

%{\it Conclusions}. 
The S–E solution describes an exact, asymptotically flat black hole supported by a dilaton field, featuring a regular event horizon and with all curvature singularities fully hidden behind the horizon. Furthermore, the spacetime obeys the null energy condition in the exterior region. Nonetheless, chronology is violated in the domain of outer communication: prior to crossing the event horizon, one encounters an open region where $g_{\phi\phi}<0$, rendering the axial orbits CTCs. Thus, even though there are no naked singularities, the exterior spacetime is causally pathological and fails to be globally hyperbolic. In this specific sense, the solution evades the usual expectation of exterior predictability (strong asymptotic predictability); that is, it violates \emph{causal/chronology censorship} rather than naked-singularity censorship.

%The S–E configuration represents a black hole that violates Penrose’s cosmic censorship conjecture, because it has been demonstrated that its event horizon lies inside the region where causality breaks down, in other words, one encounters chronological violations and the emergence of CTCs before reaching the event horizon. 
In contrast, the SU–E configuration describes a wormhole with a ring singularity that is not part of the physical spacetime, that is, it can only be obtained through the compactification of the compact object. Similarly, this wormhole has a throat at \(x_G = 0\) and \(|y| > y_1\), but for \(|y| < y_1\) the throat shifts to positions with \(x_G \neq 0\).

Figure \ref{fig:EstructuraCausalDiagrama} shows the Carter–Penrose diagram of the wormhole on the equatorial plane $y_0 = 0$. The identification of the $\pm R_G$ curves is analogous to the topological identification of a pseudo-sphere $\mathbb{S}^2$ with radius $r = \pm l_1$, and the corresponding Carter–Penrose diagram of each universe can be seen in Figure \ref{fig:DiagPen Separados}.The graphs indicates that the ring singularity and CTCs lies within the throat region of the wormhole. The identification discussed suggests that a naked singularity can be enveloped by a 'throat', analogous to how an event horizon encases a black hole's singularity, as seen in the S-E scenario. In this context, the anomalies are 'dressed' by a throat, successfully obstructing access to the ring singularity or the causal violation region. 

In summary, and most crucially, we have provided a rigorous formulation of the WCCC. The overall result is an operative and mathematically robust criterion for physically acceptable wormholes: a realistic two-universe wormhole must contain a regular manifold region that is attached to both asymptotic ends, and its throat, understood as the common causal boundary of these two ends must surround every defect set. This ensures that defects can neither be reached from, nor observed at, $\mathscr I_i^+$ in either universe. In this way, the WCCC furnishes a structural condition that separates wormholes with a truly regular exterior geometry from those setups in which pathologies exist in the vicinity but are not causally or topologically isolated.

%------------------------------------------------------------

%------------------------------------------------------------
%\vspace{-6pt}
%\section{Acknowledgements}

Acknowledgements. We thank Dr. Francisco S.N. Lobo for his enriching discussions, which were essential to developing the central idea of this work. LB thanks SECIHTI-M\'exico for the doctoral grant.
This work was partially supported by SECIHTI M\'exico under grants CBF-2025-G-1720 and CBF-2025-G-176. Also  by the grant I0101/131/07 C-234/07 of the Instituto Avanzado de Cosmolog\'ia (IAC) collaboration (http://www.iac.edu.mx/), and for the computing time granted by LANCAD and SECIHTI in the Supercomputer Hybrid Cluster "Xiuhcoatl" at GENERAL COORDINATION OF INFORMATION AND COMMUNICATIONS TECHNOLOGIES (CGSTIC) of CINVESTAV.
URL: http://clusterhibrido.cinvestav.mx/

%------------------------------------------------------------

\bibliographystyle{elsarticle-harv} 
\bibliography{Bibliografia}

@article{DelAguila:2018gni,
    author = "Del \'Aguila, Juan Carlos and Matos, Tonatiuh",
    title = "{Wormhole Cosmic Censorship: An Analytical Proof}",
    eprint = "1806.03747",
    archivePrefix = "arXiv",
    primaryClass = "gr-qc",
    doi = "10.1088/1361-6382/aaf336",
    journal = "Class. Quant. Grav.",
    volume = "36",
    number = "1",
    pages = "015018",
    year = "2019"
}

@article{Matos:2012gj,
    author = "Matos, Tonatiuh and Urena-Lopez, L. Arturo and Miranda, Galaxia",
    title = "{Wormhole Cosmic Censorship}",
    eprint = "1203.4801",
    archivePrefix = "arXiv",
    primaryClass = "gr-qc",
    doi = "10.1007/s10714-016-2040-7",
    journal = "Gen. Rel. Grav.",
    volume = "48",
    number = "5",
    pages = "61",
    year = "2016"
}

@article{DelAguila:2015isj,
    author = "Del \'Aguila, Juan Carlos and Matos, Tonatiuh and Miranda, Galaxia",
    title = "{Exact Rotating Magnetic Traversable Wormholes satisfying the Energy Conditions}",
    eprint = "1507.02348",
    archivePrefix = "arXiv",
    primaryClass = "gr-qc",
    doi = "10.1103/PhysRevD.99.124045",
    journal = "Phys. Rev. D",
    volume = "99",
    number = "12",
    pages = "124045",
    year = "2019"
}

@article{Bixano:2025jwm,
    author = "Bixano, Leonel and Matos, Tonatiuh",
    title = "{Einstein-Maxwell-dilaton wormholes that meet the energy conditions}",
    eprint = "2502.07206",
    archivePrefix = "arXiv",
    primaryClass = "gr-qc",
    doi = "10.1103/PhysRevD.111.084056",
    journal = "Phys. Rev. D",
    volume = "111",
    number = "8",
    pages = "084056",
    year = "2025"
}

@book{Chrusciel:2020fql,
    author = "Chrusciel, Piotr",
    title = "{Geometry of Black Holes}",
    isbn = "978-0-19-887320-4, 978-0-19-885541-5",
    publisher = "Oxford University Press",
    series = "International Series of Monographs on Physics",
    month = "4",
    year = "2023"
}

@article{Penrose:1964wq,
    author = "Penrose, Roger",
    title = "{Gravitational collapse and space-time singularities}",
    doi = "10.1103/PhysRevLett.14.57",
    journal = "Phys. Rev. Lett.",
    volume = "14",
    pages = "57--59",
    year = "1965"
}

@article{Penrose:1969pc,
    author = "Penrose, R.",
    title = "{Gravitational collapse: The role of general relativity}",
    doi = "10.1023/A:1016578408204",
    journal = "Riv. Nuovo Cim.",
    volume = "1",
    pages = "252--276",
    year = "1969"
}

@book{Hawking:1973uf,
    author = "Hawking, Stephen W. and Ellis, George F. R.",
    title = "{The Large Scale Structure of Space-Time}",
    doi = "10.1017/9781009253161",
    isbn = "978-1-009-25316-1, 978-1-009-25315-4, 978-0-521-20016-5, 978-0-521-09906-6, 978-0-511-82630-6, 978-0-521-09906-6",
    publisher = "Cambridge University Press",
    series = "Cambridge Monographs on Mathematical Physics",
    month = "2",
    year = "2023"
}

@article{Hawking:1966sx,
    author = "Hawking, Stephen",
    title = "{The Occurrence of singularities in cosmology}",
    doi = "10.1098/rspa.1966.0221",
    journal = "Proc. Roy. Soc. Lond. A",
    volume = "294",
    pages = "511--521",
    year = "1966"
}

@article{Hawking:1966jv,
    author = "Hawking, Stephen",
    title = "{The Occurrence of singularities in cosmology. II}",
    doi = "10.1098/rspa.1966.0255",
    journal = "Proc. Roy. Soc. Lond. A",
    volume = "295",
    pages = "490--493",
    year = "1966"
}

@article{Hawking:1967ju,
    author = "Hawking, Stephen",
    title = "{The occurrence of singularities in cosmology. III. Causality and singularities}",
    doi = "10.1098/rspa.1967.0164",
    journal = "Proc. Roy. Soc. Lond. A",
    volume = "300",
    pages = "187--201",
    year = "1967"
}

@article{Eardley:1978tr,
    author = "Eardley, Douglas M. and Smarr, Larry",
    title = "{Time function in numerical relativity. Marginally bound dust collapse}",
    doi = "10.1103/PhysRevD.19.2239",
    journal = "Phys. Rev. D",
    volume = "19",
    pages = "2239--2259",
    year = "1979"
}

@article{Shapiro:1991zza,
    author = "Shapiro, Stuart L. and Teukolsky, Saul A.",
    title = "{Formation of naked singularities: The violation of cosmic censorship}",
    doi = "10.1103/PhysRevLett.66.994",
    journal = "Phys. Rev. Lett.",
    volume = "66",
    pages = "994--997",
    year = "1991"
}

@article{Joshi:2011qq,
    author = "Joshi, Pankaj S. and Malafarina, Daniele and Saraykar, Ravindra V.",
    title = "{Genericity aspects in gravitational collapse to black holes and naked singularities}",
    eprint = "1107.3749",
    archivePrefix = "arXiv",
    primaryClass = "gr-qc",
    doi = "10.1142/S0218271812500666",
    journal = "Int. J. Mod. Phys. D",
    volume = "21",
    pages = "1250066",
    year = "2012"
}

@article{Joshi:2000fk,
    author = "Joshi, Pankaj S.",
    editor = "Bharadwaj, S. and Dadhich, N. K. and Kar, S.",
    title = "{Gravitational collapse: The Story so far}",
    eprint = "gr-qc/0006101",
    archivePrefix = "arXiv",
    doi = "10.1007/s12043-000-0164-4",
    journal = "Pramana",
    volume = "55",
    pages = "529--544",
    year = "2000"
}

@article{Miranda:2013gqa,
    author = "Miranda, Galaxia and Matos, Tonatiuh and Garcia, Nadiezhda Montelongo",
    title = "{Kerr-Like Phantom Wormhole}",
    eprint = "1303.2410",
    archivePrefix = "arXiv",
    primaryClass = "gr-qc",
    doi = "10.1007/s10714-013-1613-y",
    journal = "Gen. Rel. Grav.",
    volume = "46",
    pages = "1613",
    year = "2014"
}

@article{Bixano:2025bio,
    author = "Bixano, Leonel and Matos, Tonatiuh",
    title = "{On the Possibility of the Existence of Wormholes in Nature}",
    eprint = "2505.20167",
    archivePrefix = "arXiv",
    primaryClass = "gr-qc",
    month = "5",
    year = "2025"
}

@Article{axioms14110831,
AUTHOR = {Bixano, Leonel and Sarmiento-Alvarado, I. A. and Matos, Tonatiuh},
TITLE = {A Brief Review of Wormhole Cosmic Censorship},
JOURNAL = {Axioms},
VOLUME = {14},
YEAR = {2025},
NUMBER = {11},
ARTICLE-NUMBER = {831},
URL = {https://www.mdpi.com/2075-1680/14/11/831},
ISSN = {2075-1680},
DOI = {10.3390/axioms14110831}
}

@article{Komar:1958wp,
    author = "Komar, Arthur",
    title = "{Covariant conservation laws in general relativity}",
    doi = "10.1103/PhysRev.113.934",
    journal = "Phys. Rev.",
    volume = "113",
    pages = "934--936",
    year = "1959"
}

@article{Nedkova:2011hx,
    author = "Nedkova, Petya G. and Yazadjiev, Stoytcho S.",
    title = "{On the Thermodynamics of 5D Black Holes on ALF Gravitational Instantons}",
    eprint = "1109.2838",
    archivePrefix = "arXiv",
    primaryClass = "hep-th",
    doi = "10.1103/PhysRevD.84.124040",
    journal = "Phys. Rev. D",
    volume = "84",
    pages = "124040",
    year = "2011"
}

@article{Clement:2015aka,
    author = "Cl{\'e}ment, G{\'e}rard and Gal'tsov, Dmitri and Guenouche, Mourad",
    title = "{NUT wormholes}",
    eprint = "1509.07854",
    archivePrefix = "arXiv",
    primaryClass = "hep-th",
    reportNumber = "LAPTH-047-15",
    doi = "10.1103/PhysRevD.93.024048",
    journal = "Phys. Rev. D",
    volume = "93",
    number = "2",
    pages = "024048",
    year = "2016"
}

@article{Clement:2022pjr,
    author = "Cl{\'e}ment, G{\'e}rard and Gal'tsov, Dmitri",
    title = "{Rotating traversable wormholes in Einstein-Maxwell theory}",
    eprint = "2210.08913",
    archivePrefix = "arXiv",
    primaryClass = "gr-qc",
    reportNumber = "LAPTH-064/22",
    doi = "10.1016/j.physletb.2023.137677",
    journal = "Phys. Lett. B",
    volume = "838",
    pages = "137677",
    year = "2023"
}

@article{BallonBordo:2019vrn,
    author = "Ballon Bordo, Alvaro and Gray, Finnian and Hennigar, Robie A. and Kubiz{\v{n}}{\'a}k, David",
    title = "{The First Law for Rotating NUTs}",
    eprint = "1905.06350",
    archivePrefix = "arXiv",
    primaryClass = "hep-th",
    doi = "10.1016/j.physletb.2019.134972",
    journal = "Phys. Lett. B",
    volume = "798",
    pages = "134972",
    year = "2019"
}

@article{Misner:1963fr,
    author = "Misner, Charles W.",
    title = "{The Flatter regions of Newman, Unti and Tamburino's generalized Schwarzschild space}",
    doi = "10.1063/1.1704019",
    journal = "J. Math. Phys.",
    volume = "4",
    pages = "924--938",
    year = "1963"
}

@article{Manko:2005nm,
    author = "Manko, V. S. and Ruiz, E.",
    title = "{Physical interpretation of NUT solution}",
    eprint = "gr-qc/0505001",
    archivePrefix = "arXiv",
    doi = "10.1088/0264-9381/22/17/014",
    journal = "Class. Quant. Grav.",
    volume = "22",
    pages = "3555--3560",
    year = "2005"
}

\end{document}